\documentclass[12pt]{article}
\pdfoutput=1

\usepackage{pstricks}
\usepackage{color}
\usepackage{latexsym,amsmath,amssymb}
\usepackage{graphicx}
\usepackage{slashed}  
\usepackage{bm}
\usepackage{epsfig}
\usepackage{dcolumn}
\usepackage{cite}
\usepackage{cancel}
\usepackage{hyperref}

\usepackage{bm}
\usepackage{latexsym}
\usepackage{amsmath}
\numberwithin{equation}{section}

\def\ee{\end{equation}}
\def\ba{\begin{eqnarray}}
\def\ea{\end{eqnarray}}

\def\bq{\begin{quote}}
\def\eq{\end{quote}}

 at 10truept

\newcommand{\beq}{\begin{equation}}
\newcommand{\eeq}{\end{equation}}
\newcommand{\beqa}{\begin{eqnarray}}
\newcommand{\eeqa}{\end{eqnarray}}
\newcommand{\bea}{\begin{eqnarray}}
\newcommand{\eea}{\end{eqnarray}}
\newcommand{\p}{\partial}

\newcommand{\lmk}{\left(}
\newcommand{\rmk}{\right)}

\newcommand{\hf}{\frac{1}{2}}

\def\lesssim{~\mbox{\raisebox{-.6ex}{$\stackrel{<}{\sim}$}}~}

\def\ltap{\ \raise.3ex\hbox{$<$\kern-.75em\lower1ex\hbox{$\sim$}}\ }
\def\gtap{\ \raise.3ex\hbox{$>$\kern-.75em\lower1ex\hbox{$\sim$}}\ }
\def\gl{\ \raise.5ex\hbox{$>$}\kern-.8em\lower.5ex\hbox{$<$}\ }
\def\roughly#1{\raise.3ex\hbox{$#1$\kern-.75em\lower1ex\hbox{$\sim$}}}

\def\baq{\begin{eqnarray}}
\def\eaq{\end{eqnarray}}

\begin{document}

\begin{titlepage}

\nopagebreak
\title{  \begin{center}\bf Radiative Corrections from Heavy Fast-Roll Fields during Inflation \end{center} }

\vfill
\author{Rajeev Kumar Jain, McCullen Sandora, and Martin S. Sloth}

\date{ }

\maketitle
\thispagestyle{empty}
\begin{center}
\vspace{-0.7cm}
{\it  CP$^3$-Origins, Center for Cosmology and Particle Physics Phenomenology}\\
{\it  University of Southern Denmark, Campusvej 55, 5230 Odense M, Denmark}

\end{center}
\vfill
\begin{abstract}
We investigate radiative corrections to the inflaton potential from heavy fields undergoing a fast-roll phase transition. We find that a logarithmic one-loop correction to the inflaton potential involving this field can induce a temporary running of the spectral index. The induced running can be a short burst of strong running, which may be related to the observed anomalies on large scales in the cosmic microwave spectrum, or extend over many e-folds, sustaining an effectively constant running to be searched for in the future.  We implement this in a general class of models, where effects are mediated through a heavy messenger field sitting in its minimum.  Interestingly, within the present framework it is a generic outcome that a large running implies a small field model with a vanishing tensor-to-scalar ratio,  circumventing the normal expectation that small field models typically lead to an unobservably small running of the spectral index.  An observable level of tensor modes can also be accommodated, but, surprisingly, this requires running to be induced by a curvaton.  If upcoming observations are consistent with a small tensor-to-scalar ratio as predicted by small field models of inflation, then the present study serves as an explicit example contrary to the general expectation that the running will be unobservable.

 \end{abstract}
\noindent
\hfill \\
\vfill

\end{titlepage}

\section{Introduction}

Over the past few decades, cosmic inflation \cite{Starobinsky:1979ty,inflation} has emerged as the dominant explanation for the dynamics of the early universe, bearing many successful predictions. But while inflation is considered to be a successful paradigm, one major challenge is to identify the right underlying model of inflation in the large landscape of viable models. In a situation where even a simple model of inflation can easily fit the observational data, there are good reasons to let oneself be guided by Occam's razor and focus on the minimal single field models. On the other hand, observers are working hard to reduce the parameter space of inflationary models with measurements of the Cosmic Microwave Background (CMB) and Large Scale Structure to an increasing degree of precision \cite{planckxvi,Hinshaw:2012aka,spt,Sievers:2013ica, Abell:2009aa,Laureijs:2011gra} . The precision is gradually reaching a level where some of the simplest single field models of inflation are getting in tension with observations \cite{planckxvi}. This could be a hint that inflation is not a minimal single field model, and we are forced by data to allow for a minimal amount of new physics in addition to the inflaton\footnote{Alternatively one might also give up the simple ansatz that the inflationary potential takes a monomial or polynomial form as in the original Starobinsky model \cite{Starobinsky:1979ty}.}.  In addition there are a few other small hints of signatures beyond the standard single field slow roll paradigm that, if verified, would point more directly to new physics at play in the early universe.  One of these is the lack of power on large scales, which could indicate a large temporary running of the spectrum of density perturbations on large scales.    

It is well known that new light degrees of freedom during inflation can significantly alter the standard picture. One of the most popular examples is the curvaton \cite{Enqvist:2001zp,Lyth:2001nq,Moroi:2001ct}\footnote{For some earlier related work, see also \cite{Mollerach:1989hu,Linde:1996gt}.}. The curvaton is a light field, which is completely passive during inflation, only to become important and steal the show after inflation by imprinting its fluctuation in the sky, leading to the observed curvature perturbations. At present data cannot discriminate between the curvaton model and f.ex. the Starobinsky model \cite{Starobinsky:1979ty}, but in the future a more precise measurement of non-Gaussianity could distinguish between the two models. A measurement of $f_{NL} \sim \mathcal{O}(1)$ would rule out the Starobinsky model, while agreeing with the predictions of the curvaton model \cite{Bartolo:2003jx}.

Another logical (although less studied) possibility is to similarly consider the effect of heavy fields during inflation. It is natural to ask how heavy fields present during inflation could imprint themselves in cosmological observables \cite{Langlois:2004px,Battefeld:2014aea,Cespedes:2013rda,Noumi:2012vr,Gao:2012uq,Achucarro:2012yr,Achucarro:2012sm,Cespedes:2012hu,Jackson:2010cw}. Below we will try to answer this question by discussing in detail the possibility that radiative corrections to the inflaton potential from heavy fast-rolling fields can lead to an observable running of the spectral  index. A constant running of the spectral index is an interesting observable, which can be measured to very high accuracy in the future \cite{future}, while a short temporary burst of strong running could be related to the large scale anomalies in the CMB.

One important point in the discussion of running is to discriminate between small field models ($\Delta\phi\lesssim M_p$) and large field models ($\Delta\phi\gtrsim M_p$). While large field models with an observable level of running exist, running is generically negligible in small field models, since there is little field variation to facilitate scale dependence of parameters. Therefore, if no primordial tensors modes are discovered and we are guided by data towards small field models, a non-vanishing running of the spectral index is a strong indication of non-trivial new physics in the inflationary sector. It also makes us ask the more precise question: what kind of new physics could lead to an observable running in a small field model?  

In section \ref{sec2}, we first take a step back, and discuss how a running spectral index can be thought of in a general way in terms of {\it running by proxy}. A non-trivial result is that running by proxy requires a small field model of inflation. Thus ``running by proxy" appears as a natural way to think of running in small field models.  In a small field model the potential can generically be expanded as a Taylor series around the initial field value.  Thus, a generic potential with proxy running will be of the form
\beq
V(\phi)=V_0-M^3\phi-(m^2+\Delta m^2N)\phi^2/2+...
\eeq
where we have included the term $\Delta m^2N$ (normalized such that $N=0$ at the pivot scale) with an explicit time dependence in the potential, which in general is thought to be coming from the time dependent vacuum expectation value (VEV) of another field, which we refer to as the proxy field. That the proxy term produces negative running corresponds to a decreasing mass  ({\it wilting potential}) mimicking the same choice of sign of the third derivative of the potential that is usually required for negative running. In comparison with the generic small field model expanded to second order in the inflaton, there is only one new parameter, $\Delta m^2$, which is the minimum required for the description of running, and  completely determines its amplitude. If the running is short and related to the large scale anomalies in the CMB, a second new parameter, which determines the end of the running, also needs to be added. The ratio of the running to the tilt at the pivot scale is
\beq
\frac{\alpha_s}{n_s-1}\sim\frac{\Delta m^2}{m^2}
\eeq
The current central value for this observable is $\sim1/10$, but consistent with 0 to within $1\sigma$ \cite{fine}.  As we shall see, another surprising conclusion is that the only way to have running by proxy and an observable level of primordial tensor modes (or a large field model), it is if the proxy running is in the curvaton sector and the spectral running is induced by a curvaton.

In section \ref{CW}, we discuss how running by proxy is realized in a detailed microphysical scenario as radiative corrections to the inflaton potential from heavy fields of the Coleman-Weinberg form. We provide a simple scenario in which the parameters of the inflationary potential effectively depend on scale explicitly, inherited from the time dependence of heavy fields that are otherwise very weakly coupled to the inflaton.  Our mechanism requires the presence of at least two additional fields, one in a fast roll phase and another mediator field that is needed to maintain distance from the inflaton sector.  While the fast roll field does not couple to the inflaton in the tree level potential, one loop effects induce a logarithmic coupling, leading to a spectral tilt that varies exactly linearly in the number of e-folds.  It suffices to consider potentials of the form
\beq
V(\phi,\chi,\rho)=V(\phi)-\frac{\mu^2}{2}\chi^2+f(\phi)\rho+g(\chi)\rho^2\label{potentia}
\eeq 
where $f$ and $g$ can be chosen to be simple $\it{renormalizable}$ functions, to arrive at the desired form of the one loop potential without spoiling an otherwise single field slow roll dynamics.  Though models of this form yield a positive running, which is disfavored by data, we explicitly write two generalizations that naturally lead to negative running, namely that the messenger fields are fermions, and by using Ratra-Peebles type potentials.  The most stringent restriction of the realization we consider is that we require the inflaton to be small field, in that its field variation is sub-planckian.

The structure of the paper is therefore as follows. In section \ref{sec2} we discuss general aspect of proxy running. In particular we show why it implies a small field model with a linear term unless the proxy running is in a curvaton sector. In section 3, we show how proxy running can be realized at the one-loop level by having a heavy field undergoing a fast roll phase transition. We also discuss generalizations including a curvaton and the stability of the model to perturbations and higher order loop corrections. Finally in section \ref{concl} we conclude with a discussion of further generalizations, additional observables and future prospects of proxy running in a wilting Coleman-Weinberg type potential.

Throughout this paper, we work in natural units with $\hbar=c=1$, and the reduced Planck mass $M_{p}^2\equiv 1/8 \pi G$ is set to unity unless explicitly written.
 
\section{Running in single field models}\label{sec2}

Despite the rapid progress in measuring the CMB perturbations on smaller and smaller scales, the current observational bounds on the running of the spectral index are still relatively weak. In single field inflation the natural size of the running is $\alpha_s \sim r(n_s-1)/2$, and therefore current constraints $\alpha_s =-0.0031\pm 0.0074$\cite{fine} are only on the border of probing typical large field models. It is therefore one of the important goals for future experiments to improve the constraints on the running of the spectral index with at least an order of magnitude or two. In this experimental situation, we find it important to understand what an anomalous constant running (\emph{jogging}) could teach us about new physics, particularly in small field models of inflation, having small tensor-to-scalar ratio.

In addition we would like to emphasise that the running does not need to be constant over all the observational scales. In fact, a shorter running is more natural in many models. A short burst of fast running on the largest scales (\emph{sprinting}), could lead to a power suppression on large scales in the perturbation spectrum, and might therefore be related to the large scale anomalies in the CMB. 

Below we will first focus our discussion on jogging, and then comment on the sprinting scenario in the conclusions.

\subsection{The challenges of jogging}
\begin{figure*}[t]
\centering
\includegraphics[height=6.25cm]{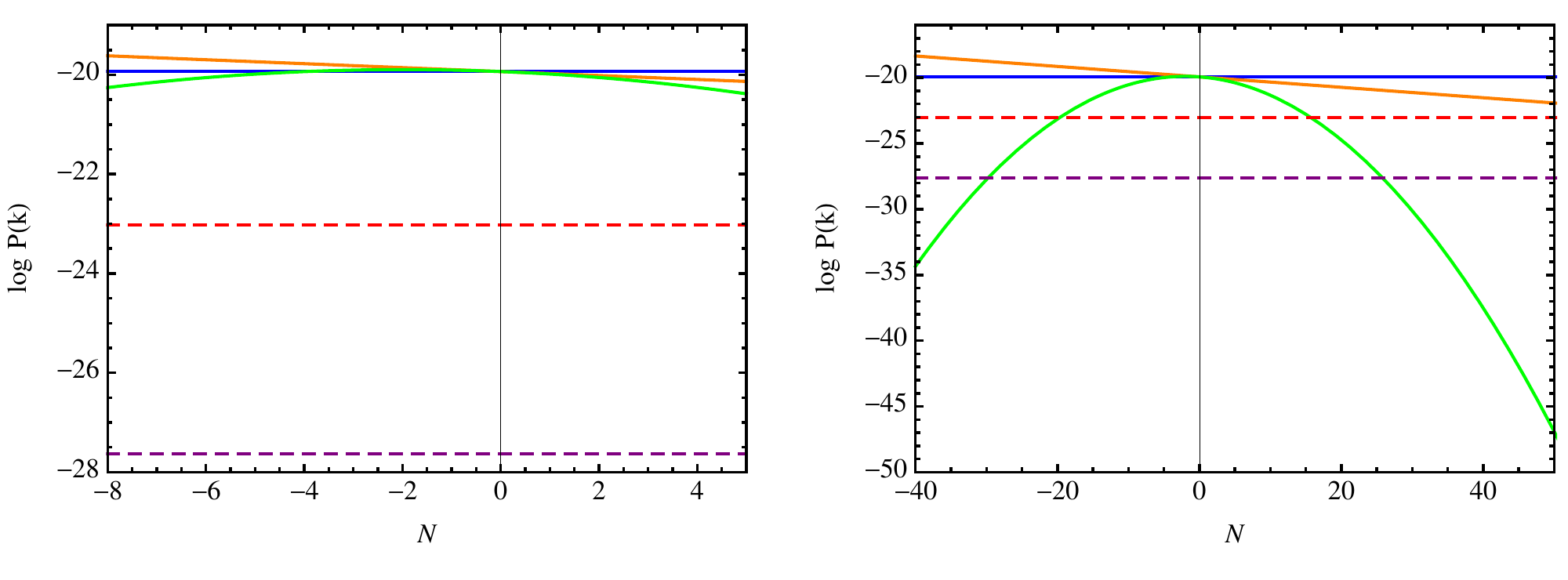}\
\caption{The power spectrum of inflationary perturbations, plotted as a function of number of e-folds (to scale).  The figure on the left shows the effects of adding a constant spectral tilt (orange) and negative running (green) to an otherwise flat spectrum (blue) within the observational window.  The figure on the right indicates the severe implications for much larger and much smaller scales.  The dashed lines represent values of $P(k)$ for which the slow roll condition breaks down, for representative values of the Hubble parameter during inflation ($H=10^{-5}$ top, $H=10^{-6}$ bottom)}
\label{pofk}
\end{figure*}

The primary observable from inflation is the power spectrum of density fluctuations as a function of the wavenumber $k$.  Because the early universe expansion is very nearly described by a de-Sitter phase, the power spectrum should be approximately scale invariant.  The vacuum energy does slowly decay, however, and so a fairly generic prediction of inflationary models is that there be more power on large scales compared to small scales.  The CMB measures roughly eight e-folds of inflation, so since the inflationary parameters only depend on time very weakly, it is usually adequate to model the data by just considering the first two coefficients of the Taylor expansion in the number of e-folds.  However, we can allow for a general expansion
\beq
\log{P(k)}=\log{P_0}+(n_s-1)N+\frac12\alpha_sN^2+\dots
\eeq
where the power spectrum is defined at some arbitrary pivot scale, usually taken to be $k=0.05\, {\rm Mpc}^{-1}$, and $N=\log(a)=\log(k/H)$ is the number of e-folds since the normalization scale left the horizon, taken here to be increasing in time.  This is represented in Fig. \ref{pofk}.  

There it can be seen that the zeroth order approximation leads to the same amount of power on all scales.  Observations indicate that $n_s-1<0$ at high statistical significance, leading to more power on large scales and less on small scales.  Extrapolating this trend back to large negative $N$, a regime where the power spectrum $P(k)\sim1$ is inevitably reached.  In this regime quantum fluctuations are as important as classical evolution, and most of the universe inflates forever in a period of eternal inflation \cite{self rep,station,big station,Goncharov:1987ir}.  Extrapolating forward to small scales, the threshold $P(k)\sim H^2$ is inevitably reached, corresponding to a breakdown of slow roll, signalling the latest possible time inflation must come to an end.  

The next order correction to this is to consider the quadratic terms, which either suppress or enhance power on both large and small scales over the linear case.  A negative running, as indicated by previous data sets \cite{planckxvi,spt,bayes}, and the corresponding lack of power on large scales have inspired some recent claims that having a negative running is incompatible with eternal inflation \cite{really?}, but this relies on extrapolating the quadratic truncation to arbitrarily large scales, which is not expected to be appropriate.  Negative running also implies that inflation must end much sooner than the linear estimate, with the difference becoming more drastic the lower the scale of inflation is.  In fact, \cite{<30} places a bound on the value of a large negative (constant) running by pointing out that the number of e-folds of inflation decreases to as little as $30$ with the observed value.  Given this, it is fair to say that the presence of a negative running has drastic implications on the dynamics of inflation, and cannot survive for any significant duration.  Conversely, positive running serves to prolong inflation, in drastic cases eternally, and increases power on both small and large scales.  Bounds were placed on the allowed value of positive running in \cite{pbh} by constraints on primordial black holes.

\begin{figure*}[t]
\centering
\includegraphics[height=7cm]{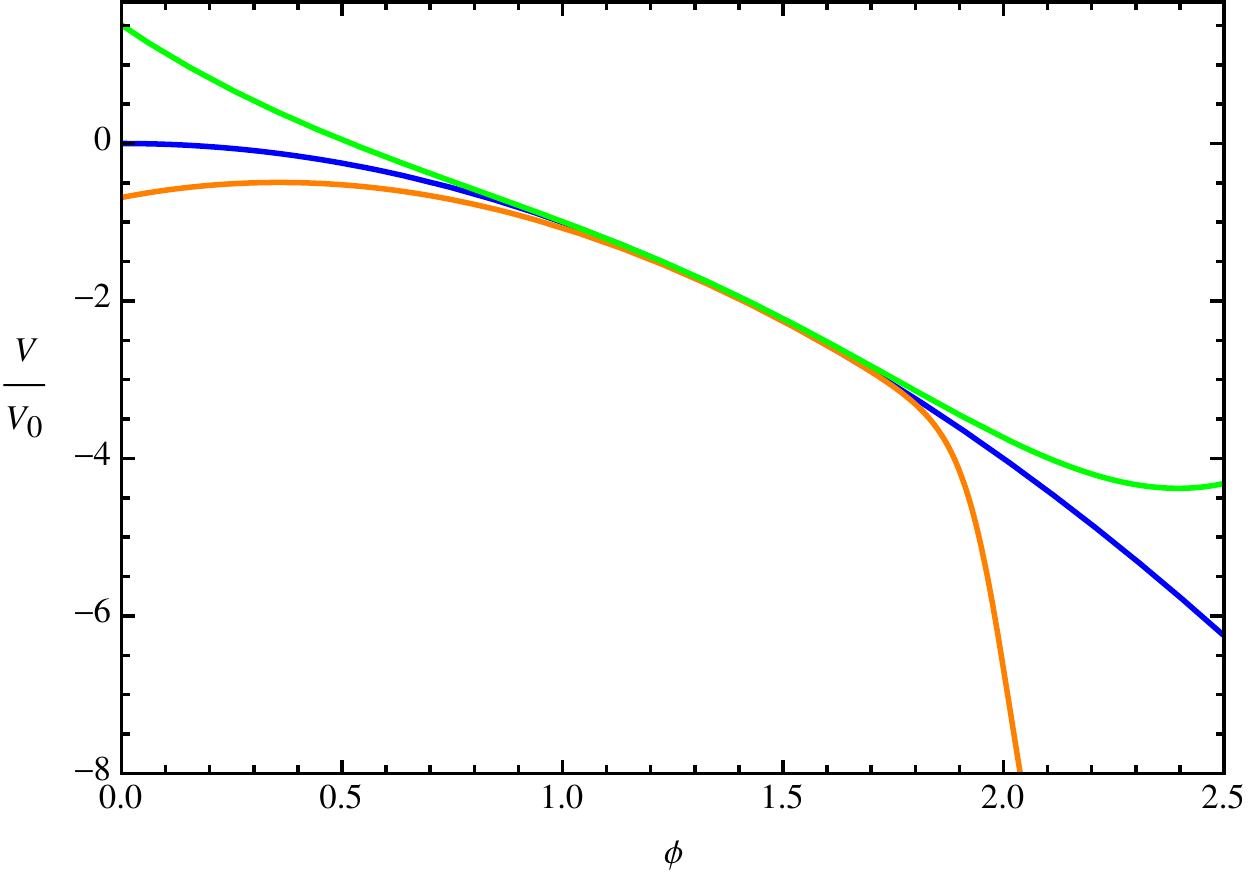}
\caption{Different schematic potentials.  The blue curve has no third derivative, which will give negligible running.  The green curve flattens out for late times, and so will give positive running.  The orange curve steepens, and will lead to negative running.}
\label{rundown}
\end{figure*}

For generic single field slow roll models, the values for $n_s$ and $\alpha_s$ are determined by parameters in the inflaton potential.  
\beq
n_s-1=-6\epsilon+2\eta,\quad \alpha_s=16\epsilon\eta-24\epsilon^2-2\xi
\eeq
where $\epsilon=V'^2/2V^2$, $\eta=V''/V$, and $\xi=V'''V'/V^2$.  Since the observed value of $n_s-1=-0.040\pm0.007$ \cite{planckxvi} and bounds on the strength of gravity waves place $r=16\epsilon<0.1$, we expect that $\epsilon,\eta \lesssim 0.01$, and their predicted contributions to the running would be far below near-future experimental detectability.  In this setup detectable running can only come from the $\xi$ term, which involves a large third derivative of the inflaton potential.  This is a possible scenario; indeed, by the reconstruction theorem \cite{reconstruct}, given any measured value of the inflationary parameters, it is possible to write down a potential that perfectly fits the data, however contrived it might seem.  The issue with this is that it is impossible to retain large running for a large amount of the inflationary potential, while maintaining slow roll.  This can be seen if we consider 
\beq
\xi\sim\mathcal{O}(1)\epsilon,\eta
\eeq
which translates into
\beq
V'''\sim \mathcal{O}(1)V',\quad V'''V'\sim \mathcal{O}(1)VV''
\eeq
If we want these relations to be generic along the inflationary trajectory, we can treat the $\mathcal{O}(1)$ coefficients as constant, or at the most very slowly varying functions.  In this case, the type of potentials that satisfy these relations are exponentials (or sinusoidal) with $V\sim V_0\,e^{\mathcal{O}(1)\phi}$, which yield slow roll parameters of order unity, and do not lead to inflation.  For large running to be present at CMB scales without destroying slow roll elsewhere, there must be a coincidence of the inflationary model.
It may be interesting to point out which type of potentials will lead to negative running.  In particular, we see that if the quantity $V'''V'$ is positive, then the running will be negative, and vice-versa, as shown in Fig. \ref{rundown}.  

The measured (constrained) values of $n_s$ and $r$ also give a subdominant contribution to the running of the form
\beq
\alpha_s= \frac12r\left(n_s-1+\frac{3}{16}r\right)+\dots\label{runninggen}
\eeq
which for all models with $r<0.02$ will be negative, but negligible, even for future planned experiments\cite{future,probe21}.  Since both this and the parameter $\xi$ are proportional to the tensor-to-scalar ratio, we generically predict that running will be utterly negligible for small field models.  This can be understood as a consequence of the small field variation in these models, which translates into the near constancy of observable parameters.

There are a number of models in the literature that achieve large running:  Instep inflation \cite{instep} uses non-polynomial and non-renormalizable potentials to provide a large third derivative term that can drive running.  Generic polynomials \cite{bigsmall} can allow for a brief period of running, but requires a fine tuning of the parameters in the potential and can only last for a few e-folds.  Modulated potentials in axionic models \cite{multisin} give dynamics where both the tilt and the running oscillate about zero.  If the modulations vary slowly over the observable window it can even look constant, and cure a slight deficit in the matter power spectrum \cite{just1}.  Note that this is necessarily a large field model, which makes its predictions for the tensor-to-scalar ratio very distinguishable from the minimal one we consider in this paper.  A falsely fat curvaton \cite{falsefat} can provide running, with a tuning of the parameters in the model, though as it is currently formulated the running is necessarily positive.

 A rather simple option was explored in \cite{curv}, where the parameters of the inflaton potential effectively become functions of the number of e-folds during inflation.  If we are after a running, but not a running of the running, then we should expect the dependence to be primarily linear.  As stated in \cite{curv}, this can be achieved if there are additional heavy fields in a phase of fast roll, where the field has a negative mass mildly larger than the Hubble rate.  We detail this scenario, which can ultimately be achieved using only renormalizable couplings, below.

\subsection{Proxy running}\label{proxrun}
In what follows we consider inflation with a general potential $V(\phi,N)$, which explicitly depends on the field $\phi$ as well as the number of e-folds $N$.  We denote derivatives with respect to $\phi$ as primes, and partial derivatives with respect to $N$ as subscripts.  With this notation, the expression for the spectral tilt is
\beq
n_s-1=-6\epsilon+2\eta+3\frac{V_N}{V}-2\frac{V'_N}{V'}-2\Gamma_0\frac{V_N''}{V}
\eeq
The first four terms result from the changes in the amplitude of fluctuations, $P(k)\propto V^{3}/V'^2$, while the last term comes from a change in the normalization, as pointed out for the case of constant parameters in \cite{accu}.  The constant $\Gamma_0=2-\log(2)-\gamma_E\approx 0.73$.  The running is
\begin{eqnarray}
\alpha_s=16\epsilon\eta-24\epsilon^2-2\xi+4\frac{V''_N}{V}+(18\epsilon-2\eta)\frac{V_N}{V}
-(18\epsilon+2\eta)\frac{V'_N}{V'}-3\frac{V_N^2}{V^2}+2\frac{{V'_N}^2}{V'^2}\nonumber\\
+\,3\frac{V_{NN}}{V}-2\frac{V'_{NN}}{V'}
-2\Gamma_0\left(\frac{V''_{NN}}{V}-\frac{V_NV''_N}{V^2}\right)+(2+12\Gamma_0)\epsilon\frac{V''_N}{V}+\frac43\eta\frac{V''_N}{V}
\end{eqnarray}

A number of general conclusions can be made from these two expressions.  First, because the explicit e-fold dependence gives a contribution to the tilt as well as the running, we demand that $V_N/V, V'_N/V'\lesssim\mathcal{O}(1/100)$, the same order as the slow roll parameters.  These terms may in fact dominate the expression for $n_s$, giving a tilt to otherwise completely flat models, but these additional terms should not be much larger than that.  Then, virtually every term in the running is second order in slow roll. Exceptions include the first two terms on the last line, which we take to be zero to avoid running of the running\footnote{and in fact is predicted by the models we consider in this paper}, and the fourth term.  Thus we have $\alpha_s\sim4V''_N/V$.

We note that in order for the contribution of the proxy e-fold dependence to $n_s$ to not greatly exceed the contribution to $\alpha_s$, we require $V_N/V\lesssim V''_N/V$.  For simple polynomial potentials we have $V''_N\sim V_N/\phi^2$, and we immediately see that the contribution to the tilt will be much larger than the running in large field models, where $\phi>1$.  From this we conclude that the running by proxy scenario is better suited for small field inflation.  Of course, our conclusion may be circumvented by considering non-polynomial potentials, but this is exactly the situation we are trying to avoid by invoking explicit time dependence in a renormalizable potential.  One scenario that naturally circumvents this conclusion is the curvaton, which will be discussed in section \ref{curvaton}. 

A final remark severely constrains the setting in which this mechanism can occur:  we first rewrite the dominant expression for the running as $\alpha_s=4V''_N/V''\eta$, where we now want $V''_N/V''\sim\mathcal{O}(1)$ in order for the running to be large.  Again, for polynomial potentials, though, we generically have $V''_N/V''\sim V'_N/V'$, which we constrained to be of order the slow roll parameters by demanding a nearly flat spectrum.  It appears as though this precludes our mechanism from being operational.  There is an important caveat, however, for which this estimate is incorrect:  If inflation is driven by a linear term, then its second derivative is zero to leading order, and the contribution to the running can be hierarchically larger than the tilt.  This is actually a very generic situation, as in small field models the potential should usually be thought of as an effective Taylor expansion around some point.  The leading order term will be a constant energy density, but the next to leading term is linear.  The details of this model will be borne out below.

One potential concern is that if we expect a model to have large running, then the running of the running would also be large.  We will see that contrary to these general expections, the Coleman-Weinberg setting has the definite prediction that the running be large, with quadratic dependence on $N$ suppressed.

\subsection{Inflating on a wilting potential}
\begin{figure*}[thb]
\centering
\includegraphics[height=7cm]{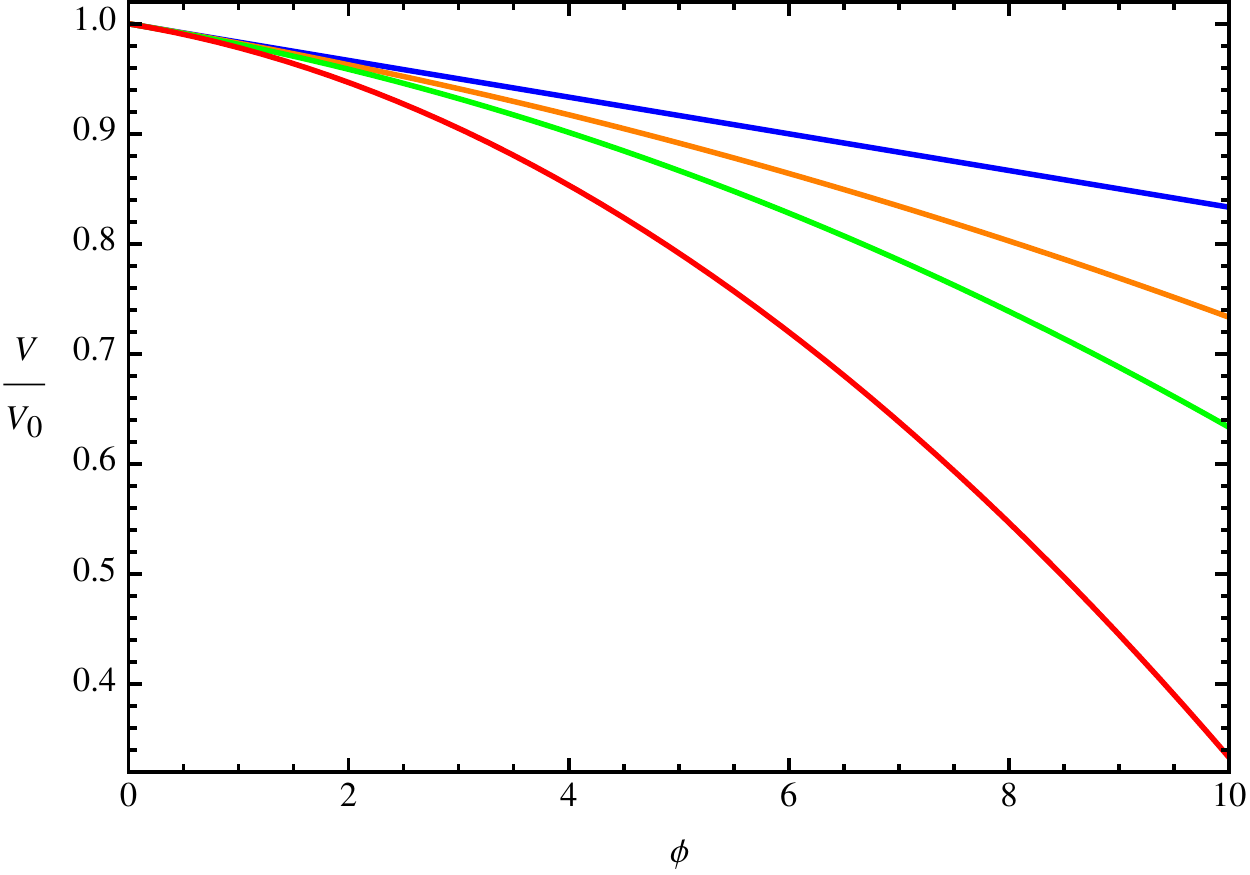}
\caption{The effective potential at several different times. As the number of e-folds increases, the potential becomes more and more negatively curved, resembling a wilting leaf or flower.}
\label{wilt}
\end{figure*}

We consider a generic small field inflation potential with explicit dependence on the number of e-folds, to demonstrate the ease with which we get large running.  Realizations of this model from loop effects is the subject of section \ref{CW}.  Because we are considering small field models, the potential can generically be expanded as a Taylor series around the initial field value.  Thus, a generic potential will be of the form
\beq
V(\phi)=V_0-M^3\phi-(m^2+\Delta m^2N)\phi^2/2+...
\eeq
where we take the higher order terms to be small, both on the grounds that our Taylor expansion framework is valid, and from the observational fact that inflaton self interactions are small.  The sign of the linear term is a convention, but we have included a suggestive minus sign for the quadratic terms, which reproduces a negative running (and spectral index).  As of now this is a mere notational trick, and we may just as well think of $m^2$ as being \emph{a priori} either positive or negative, to be set by observations, and hopefully arrived at naturally in a more complete framework.  That this choice of sign produces negative running is clear:  it corresponds to a decreasing mass, which mimics the same choice of sign of the third derivative of the potential that is usually required for negative running, and resembles a wilting flower, as shown in Fig. [\ref{wilt}].  This will lead to the slow roll parameters becoming increasingly large on small scales, which will tend to end inflation early. (If running were truly constant, this makes the observed value hard to reconcile with having more than 30 e-folds of inflation \cite{<30}.  This worry is automatically avoided if the time dependence comes from integrating out a field in the fast roll phase, since the field can simply stop fast rolling at some point within a currently unobservable regime.  One prediction of this model is that once one could probe down to these scales, an abrupt halt of the running would be observed.)
This model can easily be made to fit all observations, if the following conditions are met: $V_0\gg M^3\phi\gg m^2\phi^2$.  Then the dominant contributions to the slow roll parameters are
\beq
\epsilon=\frac{M^6}{2V_0^2},\quad \eta=-\frac{m^2+\Delta m^2 N}{V_0},\quad \frac{V_N}{V}=-\frac{\Delta m^2\phi^2}{2V_0},\quad \frac{V'_N}{V'}=\frac{\Delta m^2\phi}{M^3},\quad \frac{V''_N}{V}=-\frac{\Delta m^2}{V_0}\label{slowpar}
\eeq
The power spectrum is given approximately by 
\beq
P(k)\simeq\frac{1}{12\pi^2}\frac{V_0^3}{M^6}
\eeq
which when normalized to the observed value of $2.2\times10^{-9}$ gives the condition
\beq
\frac{V_0^{1/2}}{M}\approx0.08
\eeq
The value of the spectral tilt will be 
\begin{eqnarray}
n_s-1&=&3\frac{M^6}{V_0^2}-2\frac{m^2+\Delta m^2 N}{V_0}-\frac32\frac{\Delta m^2\phi^2}{V_0}-2\frac{\Delta m^2\phi}{M^3}+2\,\Gamma_0\frac{\Delta m^2}{V_0}\nonumber\\
&\approx&-2\frac{m^2+\Delta m^2 N}{V_0}+2\,\Gamma_0\frac{\Delta m^2}{V_0}
\end{eqnarray}
The value for the running is
\beq
\alpha_s=-4\frac{\Delta m^2}{V_0}
\eeq
This is of comparable size to the tilt if the tilt's dominant component comes from the proxy term, i.e. if the mass parameter in the potential vanishes.  The ratio of the running to the tilt is
\beq
\frac{\alpha_s}{n_s-1}=\frac{2\Delta m^2}{m^2-\Gamma_0\Delta m^2}
\eeq
The current central value for this observable is $\sim1/10$, but is consistent with 0 at the $1\sigma$ level. 

Now we consider the latest possible time inflation can end is when the potential becomes negative.  If the reheating temperature is much lower than the scale of inflation, we can neglect it and state that the end of inflation occurs when $V(\phi)=0$, otherwise we can allow for a non-negligible (perturbative) reheating temperature and demand $V(\phi)=\sigma T_{rh}^4$, where $\sigma = \frac{\pi^2}{30} g_{*}$ with $g_{*}\simeq 200$ being the effective number of degrees of freedom. Inflation will then end at 
\beq
\phi_*=\frac{M^3}{m^2}\left(\sqrt{1+\frac{2m^2}{M^6}(V_0-\sigma T_{rh}^4)}-1\right)
\eeq
An approximate solution for the value of the inflaton, assuming the slow roll conditions, is 
\beq
\phi(N)=\frac{M^3}{m^2}(e^{|\eta| N}-1)\label{prophile}
\eeq
For $N\ll |\eta|^{-1}$ this reduces to $\phi(N)\approx M^3N/V_0$, so that the inflaton spends $\mathcal{O}(50)$ e-folds in the linear regime before transferring to exponential behavior.  One difference from the standard case is that the value of $\eta$ keeps growing in this scenario until the end of fast roll.  In general we do not expect fast roll to continue for much longer than the scales probed by the CMB, so the final value of $\eta$ will be roughly the same as its value at the pivot scale, $\sim1/50$.  As the field is still in the linear regime at the point where $\eta$ stops evolving, to estimate the total number of e-folds it suffices to take the final value of this parameter, and the value of $\epsilon$ from eq. (\ref{slowpar}).  The total number of e-folds is
\beq
N_*=\frac{1}{2|\eta|}\log\left[1+\frac{|\eta|}{\epsilon}\left(1-\frac{\sigma T_{rh}^4}{V_0}\right)\right]
\eeq

The only requirement is that this must be larger than the minimal number of e-folds required to solve the horizon problem, given by
\beq
N_{\rm min} \simeq 67 + \frac{1}{4} \log \left(\frac{\rho_{\rm end}}{M_p^4}\right)
\eeq
In the scenario we consider, the total number of e-folds is generically larger than  $N_{\rm min}$, but the value depends on the details of reheating. 

%

\subsection{Realizing proxy running}
When it comes to actually realizing a linear dependence on $N$ as the expectation value of some other field, there are several possible scenarios.  The simplest option would seem to have it be related to the VEV of some heavy field that depends linearly on time.  There are essentially two options for accomplishing this:  the field could get its time dependence from the fields originally present in the theory, or it could be time dependent of its own accord.  If the dependence comes from the inflaton, we are in the situation where the heavy field is in the minimum of its potential, but with a VEV that depends on $\phi$, say $V(\sigma)=M^2(\sigma-f(\phi))^2+\alpha\sigma \phi^2$. The linear coupling to the inflaton is supposed to be subdominant, but necessary to communicate the running to the inflationary sector.  In this case, if $\sigma$ actually functions as a heavy field, then it can be integrated out and completely rewritten as an effective potential for the inflaton\footnote{assuming the heavy field stays in its minimum; otherwise the state could be modified \cite{veryhigh}}.  Here we are right back where we started, at the point where we write down a particular desired potential from the infinite space of possibilities, albeit with a bit more assurance that we can arrive at our choice from reasonable microphysics.  The usual arguments that the running cannot last apply, and we conclude that this method is unlikely to yield the behavior we want.  The same considerations apply when the field VEV depends on the curvature $R$ as well, as the time dependence of $R$ only comes about through the time dependence of $\phi$.

The other option is for the heavy field to naturally depend on time itself.  The problem is that heavy fields quickly move to their minimum, where they just contribute to the vacuum energy.  A setup with sustained time dependence is for the additional field to be in a period of fast roll, where the potential is concave down with a mass larger than the Hubble rate.  In this case, the field depends exponentially on time, and must couple logarithmically to the inflaton.  If the logarithm comes from a Coleman-Weinberg type one loop effective potential this naturally induces a linear dependence on $N$, with higher order corrections suppressed by additional loop factors.

\subsection{Review of fast roll}

As in \cite{fastroll, curv}, we consider a \emph{fast roll} potential for a heavy field $\chi$, given by 
\beq
V(\chi)= V_{0} - \frac{1}{2} \mu^2 \chi^2,
\eeq
where $\mu>H$. This potential leads to exponential time dependence of a homogeneous background field  due to the tachyonic instability caused by the negative mass squared term:

%
%
\beq
\chi(t) = \chi_{0}\, e^{F Ht}
\eeq 
where 
\beq
F\lmk\frac{\mu^2}{H^2}\rmk =  \sqrt{\frac{9}{4} + \frac{\mu^2}{H^2}} -\frac{3}{2}.\label{defF}
\eeq
In the limit $\mu \gg H$, one gets $F\left(\frac{\mu^2}{H^2}\right) \approx \frac{\mu}{H}$ and $\chi(t)\approx\chi_0e^{\mu t}$.  As discussed in \cite{imaginary,fastroll}, a reasonable expectation for the initial value of the field would be the mass parameter $\chi_0\sim\mu$, possibly multiplied by an $\mathcal{O}(1)$ number.  This initial value will not be important for our purposes, both because we do not expect to observe the exact beginning of fast roll, and because the initial value will only contribute a logarithmically small shift to the mass of the inflaton.

Let us admit a generic fine tuning of this type of potential: as a negative quadratic potential is unbounded from below, it is unsuitable for a fundamental theory, and must at some point be regulated by higher powers of the field, which we take here to be quartic: $\lambda\chi^4$.  Fast roll will cease when the field reaches the minimum of its potential $\chi\sim\mu/\lambda^{1/2}$.  For fast roll to last a significant number of e-folds we need
\beq
\lambda\ll e^{2FN_{\text{TOT}}}
\eeq
where for constant running we want $N_{\text{TOT}}$ to be at least as large as the observable window, which corresponds to $\lambda<(1/2500)^{2F}\sim10^{-7F}$, and if the fast roll period extends beyond that the tuning only gets worse.  This looks like a big fine tuning in the fast roll sector, especially if $\mu\gg H$, so we prefer to keep them the same order of magnitude.  We note that in standard inflationary scenarios the quartic coupling must be taken to be extremely tiny anyway, say $\lambda\sim10^{-12}$, in order to satisfy constraints from data\cite{chaotic,hilltop}, so the tuning in this model is not necessarily much worse than standard.  Finally, we should note that an extremely tiny coupling is not spoiled by quantum gravity effects, as these will give contributions of the form $\Delta\lambda\sim(H/M_p)^2$, which will be tiny for low scale inflation.

\section{Proxy running from loop effects}\label{CW}
In this section we explore obtaining running in the inflaton sector through the presence of a fast rolling field in a hidden sector.  Since the fast roll field has exponential time dependence, a linear running can only be obtained if the effective potential contains a logarithm of the fast roll field.  Thankfully, this is a generic expectation of Coleman-Weinberg type effective potentials \cite{cole}, and so we are driven to consider their use to see how this expectation may be realized.  There are several subtleties that arise in the actual implementation of this plan, so we summarize the main results here before embarking on the full calculation.

We first note that the inflaton cannot couple to the fast roll field directly for two reasons:  as mentioned above, such a coupling would induce an exponentially quick running of the spectrum, ruining slow roll in just a few e-folds.  Secondly, as we will show in detail below, the negative mass for the fast roll field prevents us from interpreting its loop effects as coming from an effective potential, essentially because the resulting effective action would be non-unitary.  Both of these considerations combined lead us to introduce a third field $\rho$ that acts as a messenger field between the inflaton and fast roll sectors.  We refer to $\rho$ as the $\it{marathon}$, because it provides the setting for a large amount of running to take place.  Generic couplings to this field will still not provide us with a 1-loop potential appropriate for running to occur, but if we take the special form
\beq
V(\phi,\chi,\rho)=V(\phi)-\frac{\mu^2}{2}\chi^2+f(\phi)\rho+g(\chi)\rho^2
\eeq
where $\phi$ couples linearly to the marathon and $\chi$ couples quadratically, we do end up in a situation with running.  The logic behind this choice can be seen if we consider the inflaton-marathon sector only, and take the VEV of the marathon to vanish:

The Hessian is then
\begin{equation*}
V_{ij}=\begin{pmatrix}
V''(\phi) & f'(\phi) \\
f'(\phi) & 2g(\chi)
\end{pmatrix}
\end{equation*}
so that the effect of the linear marathon term is to induce an off-diagonal mixing term that introduces a $\phi$ dependence into an otherwise completely sequestered scenario.  We will show that the 1-loop Coleman-Weinberg potential is of the form
\beq
V_{\text{1-loop}}=\frac{1}{64\pi^2}\sum_i(m_i^2-2H^2)^2\left(\log(m_i^2-2H^2)-\frac32\right)\label{slaw}
\eeq
where the $m_i^2$ are the eigenvalues of the Hessian.  In the limit where the off-diagonal terms are very small, the dominant eigenvalue becomes $m^2=2g+f'^2/2g$, leading to an effective potential of the form
\beq
V_{\text{1-loop}}=\frac{1}{32\pi^2}f'(\phi)^2\log g(\chi)+\dots
\eeq
and we are free to choose simple functions to reproduce the desired behavior. 

Fleshing this argument out, along with embedding this mechanism into the more complete Hessian, is the subject of this section.
We begin with some general subtleties of multifield Coleman-Weinberg potentials in section \ref{multi}, and outline the general behavior of perturbation terms needed to get running.  In section \ref{imact} we discuss the necessary complications that take place when one of the fields is a tachyon, the correct way to interpret the field dynamics, some potentially observational consequences of this setting, and finally we discuss the effects of curvature on the effective potential and show that they are small.  In section \ref{mediators} we discuss the full embedding into a three field model, and list all the constraints that must be satisfied in our scenario.  In section \ref{simple} we write down a simple example potential that leads to positive running.  The next two sections \ref{neginv} and \ref{negferm} are devoted to two different modifications on this simple scenario that lead to negative running.  Section \ref{curvaton} extends the discussion to the case where the running is in a curvaton sector.  We discuss the contribution of the perturbations of the fast roll field to the curvature perturbation in \ref{superh}, and show the effect is negligible.  We conclude in \ref{higher} with a discussion on the validity of our potential under higher loop corrections. 

\subsection{Coleman-Weinberg with multiple fields}\label{multi}
We start with a generic action of multiple fields $\phi_i$, labeled by an index $i$, in flat space-time, and then return to discuss the robustness of the derived potential to curvature corrections at the end of section \ref{imact}.  Denoting field variations by subscripts corresponding to the field's index, the expression for the 1-loop effective action 
\beq
\Gamma=-\hf \log (\text{det}\, S_{ij})\label{effective}
\eeq
yields an expression for the effective potential
\beq
V_{\text {eff}}=\frac{1}{16\pi^2}\int_0^\infty dk k^3\text{Tr}\log (k^2\delta_{ij}+V_{ij})
\eeq
which involves the Hessian matrix of the tree level potential $V$ evaluated on homogeneous backgrounds of the fields.  If this matrix is diagonalized the determinant becomes a product of its diagonal components, and the logarithm splits into a sum
\beq
V_{\text {eff}}=\frac{1}{16\pi^2}\int_0^\infty dk k^3\sum_i\log(k^2+m_i^2)\label{split}
\eeq
For the remainder of this section we take all eigenvalues to be positive, to explore the general forms that appear in the 1-loop potential.  In our general setup at the very least the eigenvalue corresponding to the fast roll field will be negative, leading to an imaginary term in the effective action, and preventing us from interpreting the field dynamics as arising from an effective potential.  The procedure to be outlined entails excising all negative eigenvalues from the sum, and handling them at the level of correlators as outlined in the next section.  Restricting to positive terms only, then, we have
\beq
V^+(\phi_i)=\frac{1}{64\pi^2}\sum_{m_i^2>0}(m_i^2)^2\left(\log(m_i^2)-\frac32\right)\label{slaw}~,
\eeq
which is stable to curvature corrections as long as $m_i^2 \gg H^2$, as we will show at the end of section \ref{imact}. We should comment here also on our choice of renormalization prescription.  As always, there will be divergent terms that must then be reabsorbed into the coupling constants in the original Lagrangian, which must ultimately be set by matching the parameters with observed quantities at some observable scale.  Because we are in the inflationary context, however, our life is easy, since the only matching comes from the single experiment of observing $P(k)$, the power spectrum of temperature fluctuations in the CMB.  Thus we are free to choose the most convenient renormalization prescription for calculations, and in the end compute the observables $P(k),n_s,r$ from the potential we arrive at.  The logarithmic dependence will then determine the running of these parameters unambiguously as a function of dynamics of the fields.  Though all the arguments of our logarithms are dimensionful quantities that must be compensated by some arbitrary scale, it is adequate for these purposes to choose the smallest threshold mass in the theory \cite{dispar}, which in this case is the mass of the inflaton.  In what follows all arguments of the logarithms are understood to be divided by the inflaton mass $m$.

We have seen that the effective potential is a sum of tems of the form $V_{+}=\sum M^4\log M^2$.  In order to get a running, we need for one of the eigenvalues to be approximately equal to some power of the fast roll field $\chi$.  It may then be tempting to take the eigenvalue to be $M^2=\chi^mf(\phi)$, as then the dependence of the inflaton in the logarithm would factor out.  However, this simple situation leads to 
\beq
V_+\approx\chi^{2m}f(\phi)^2\left(\log(\chi)+\log(f(\phi))\right)
\eeq
so that the effective inflaton couplings would run like a power of $\chi$, much too quickly for the behaviour we are after, and quickly ruining slow roll.

We are forced to consider a more complicated scenario: if the leading term is completely independent of $\phi$ then this will not contribute to the effective potential for the inflaton, but instead will just alter the dynamics of the fast roll field, which is an unobservable detail at the moment.  We would then be interested in the sub-leading terms in the eigenvalue $M^2$, which would be the leading contribution to observable quantities in the sky.  If we denote $M^2=M^2_{dom}(\chi)+M^2_{sub}(\phi,\chi)$ where $M^2_{sub}\ll M^2_{dom}$ is the largest contribution to $M^2$ that depends on $\phi$, then the leading contribution to the effective potential will be
\beq
V_+\approx M^2_{dom}M^2_{sub}\log(M^2_{dom})
\eeq
If we want the prefactor to be independent of the fast roll field, we need $M^2_{sub}(\phi,\chi)=f(\phi)/M^2_{dom}(\chi)$ so that the $\chi$ dependence will cancel.  Fortunately, this apparently bizarre tuning occurs quite naturally in these setups, as we will now demonstrate in a simple two field model.  In this case, the expression for the eigenvalues for the potential can be written explicitly as
\beq
V_{\text{1-loop}}=\frac{1}{64\pi^2}\left[(m_+^2)^2\left(\log(m_+^2)-\frac32\right)+(m^2_-)^2\left(\log(m_-^2)-\frac32\right)\right]
\eeq
where
\beq
m^2_{\pm}=\frac{V_{11}+V_{22}}{2}\pm\hf\sqrt{(V_{11}-V_{22})^2+4V_{12}^2}\label{eigem}
\eeq
From the expression for the masses we can notice some generic properties:  firstly, that they are both manifestly real, as a consequence of $V_{ij}$ being a Hermitian matrix, and second, that they can be both positive or negative.  If the diagonal elements dominate, then the off-diagonal term can be treated as a perturbation, and the masses will be approximately equal to the diagonal elements.  If the off-diagonal element is much larger than the diagonals $V_{12}^2>V_{11}V_{22}$ then one direction is guaranteed to be negative.  In this case the potential becomes very steep along the direction $\phi_1=\pm\phi_2$, and it would be energetically beneficial for one of the fields to approach zero, and the system will be dynamically driven to the regime where both eigenvalues are positive.  We stay away from this case, as it occurs only in a transient regime and does not yield the desired behavior.  If we are in the situation where $V_{12}^2\ll V_{11}V_{22}$, $V_{22}\gg V_{11}$ then the square root can be expanded:
\beq
m_+^2\approx V_{22}+\frac{V_{12}^2}{V_{22}},\quad m_-^2\approx V_{11}-\frac{V_{12}^2}{V_{22}}
\eeq
with signs of the subdominant terms enforced by the generic phenomena of level splitting, wherein small perturbations to a system generically cause the spacings between its eigenvalues to increase.
Then, the cross terms in the square of $m_+^2$ are independent of the leading behavior, 
\beq
(m_+^2)^2=V_{22}^2+2V_{12}^2+\frac{V_{12}^4}{V_{22}^2}
\eeq
This is not so mysterious:  from the characteristic equation determining the eigenvalues, we can rewrite their squares as
\beq
(m_{\pm}^2)^2=(V_{11}+V_{22})m_{\pm}^2-V_{11}V_{22}+V_{12}^2\label{splitting}
\eeq
so that the dependence on the off-diagonal element comes from the determinant of the matrix, where there is no communication with the diagonal entries.  There are also the higher order terms in the expansion of powers of $V_{12}^2/V_{22}^2$, but as long as $V_{22}$ is large enough these terms will be suppressed, which is exactly the case when $V_{22}\sim\chi^p$, in which case the higher order terms vanish exponentially.  After some comments on the correct handling of the negative eigenvalues, we show how to embed this Hessian into a realistic potential.

\subsection{Imaginary action}\label{imact}
If no fields were tachyons all eigenvalues would be non-negative, but if one of the fields is in the fast roll regime, then that direction will be tachyonic, meaning its eigenvalue will be negative.  This is a disaster for our interpretation of the one loop resummation as an effective potential, since the logarithm of a negative number is complex, yielding an effective action which is nonunitary.  The interpretation of this phenomena has been well studied in flat space, and for a single field in \cite{imaginary}, and its applications to inflation studied in \cite{inverted}, so we will consider how this works in the case of a single tachyonic field first.  There it was noted that the instability results from an improper analytic continuation of the frequencies of tachyonic modes, and represents a physical tendency for long wavelength modes to develop VEVs.  Indeed, any mode $k<|\mu|$ behaves as an inverted harmonic oscillator, and will exhibit exponential growth.  Even though this behavior cannot be captured as arising from an effective potential for the field in a background homogeneous state, $n$-point correlators can nevertheless be computed by considering the system to be in a modified state, where each long wavelength mode is initially in a minimum-uncertainty (Gaussian) wave packet.  Matching this state to the vacuum at early times gives a natural value for the width of the Gaussian, allowing for a unique statistical prediction for correlators.  Explicitly, the state will yield
\begin{equation*}
\langle\sigma_k\sigma_{-k}\rangle=g(k)=\left\{
\begin{array}{cl}
\displaystyle\frac{\cosh \left(2|\omega(k)|t\right)e^{-\frac{3H}{2}t}}{2|\omega(k)|} & \frac{k^2}{a^2}<|\mu^2+\frac94H^2|\\[6mm]
\displaystyle\frac{1}{2\,\omega(k)} & \frac{k^2}{a^2}>|\mu^2+\frac94H^2|
\end{array} \right.
\end{equation*}
where $\omega(k)=\sqrt{\frac{k^2}{a^2}-\mu^2-\frac94H^2}$.

This correlator exhibits exponential growth for wavelengths larger than the scale $\mu^{-1}$, but we must check that its effect on observational quantities is under control. It must also be taken into account when examining the equations of motion for the background field \cite{boy1,boy2}:  The correct procedure for this case is to derive the equations of motion {\it before} integrating over momentum in the effective potential.  The inverse propagator that then appears in the equations of motion can then be reinterpreted as a two point correlator of the field.
\beq
-\Box\sigma+V_\sigma+\int\frac{d^4k}{(2\pi)^4}\frac{V_{\sigma\sigma\sigma}}{k^2+V_{\sigma\sigma}}=-\Box\sigma+V_\sigma+V_{\sigma\sigma\sigma}\int\frac{d^3k}{(2\pi)^3}
\langle\sigma_k\sigma_{-k}\rangle=0
\eeq
From here we can see that the buildup of long wavelength modes serves as a source for the background field $\sigma$, altering its dynamics with nontrivial momentum dependence.  The exponential sensitivity to quantum fluctuations is washed out by taking the average quantity, and yields a statistical prediction for the amount of backreaction a given realization will have.  

This equation was arrived at using the background field approximation, and ceases to be valid if the integral becomes of the same order of magnitude as the first derivative of the potential.  When this occurs one may try to employ some more sophisticated resummation scheme, such as Hartree-Fock if $V_{\sigma\sigma\sigma}$ is sufficiently small, but generically one would be forced to resort to numerical techniques, as the system will be fully in the nonperturbative regime of domain growth\cite{preheat}.  However, this condition is met only once the fast roll regime comes to an end, when the field reaches a local minimum of its potential that regulates the runaway tachyonic behavior.  In this paper we want the field to remain in fast roll for the entire observable window, which means  that the integral can essentially be neglected and the background dynamics caused only by the slope of the potential.

Now we extend these considerations to a two field system consisting of the inflaton and a fast roll field.  This setup is still not complete enough to give a large running to the power spectrum, but once these effects are combined with the eigenvalue splitting method of the previous section we will be able to write explicit models that can give rise to running.

The equations of motion for the fields will then become
\ba \label{mix}
-\Box\phi+V^+_\phi+(m^2_-)_\phi\int\frac{d^4k}{(2\pi)^4}\frac{1}{k^2+m^2_-}=0\nonumber\\
-\Box\chi+V^+_\chi+(m^2_-)_\chi\int\frac{d^4k}{(2\pi)^4}\frac{1}{k^2+m^2_-}=0
\ea
Here the $1/(k^2+m^2_-)$ can be interpreted as the Green's function for a field $\sigma_-$, which is a linear combination of the fields $\phi,\chi$ that diagonalizes the potential matrix.  This can then be replaced by the function $g(k)$.

Therefore the background values of both the $\phi$ and $\chi$ fields respond to the buildup of large wavelength tachyonic modes, which in a statistical sense act as a shift in the background values of the fields' zero modes.  Because the Hessian matrix is not diagonal, the $\phi$ field does not exactly correspond to the positive eigenvalue direction, but instead is some mixture of the exponentially growing negative eigenvalue direction as seen from (\ref{mix}). In this paper we will however assume for simplicity that $(m^2_-)_\phi /V^+_\phi \ll \left< \sigma_-^2\right>$ and $(m^2_-)_\chi/V^+_\chi \ll \left< \sigma_-^2\right>$, such that the mixing is suppressed by the ratio of off-diagonal to diagonal elements and can be ignored in the background equations of motion for the inflaton and the proxy field. We plan to further consider the regimes were these conditions break down
 in future work.

One may have also noted that in the tree level Lagrangian we have allowed for the inflaton to have a negative bare mass, an allowance that may seem worrisome given the care we took to excise the fast roll field from the loops.  However, the fact that the inflaton mass is less than the Hubble parameter makes this case qualitatively different from a fast roll field, and actually afflicts any light field regardless of the sign of its mass.  This is because the effective potential we have calculated was based in flat space, whereas in an expanding spacetime any field gets a negative mass shift.

This can be seen by replacing the flat space propagator with the de-Sitter version in (\ref{effective}).  By rescaling the fields by one power of the scale factor and using conformal time as well as comoving momentum, the Lagrangian for the fields is identical to that of flat space, except that the mass becomes a time-dependent function multiplied by two powers of the scale factor and shifted by the scalar curvature $m^2 \to a^2(m^2 -2H^2)$. Switching back to physical momentum to perform the integral, all dependence on the scale factor factors out.  Following the previous flat space approach with the shifted mass, one arrives at the final form of the effective potential in de-Sitter space:
\beq
V_{eff}=-\frac{1}{64\pi^2}\left(m^2-2H^2\right)^2\left(\log\left(m^2-2H^2\right)-\frac32\right)~,
\eeq
where we ignored all $UV$ divergent terms, which are regularisation scheme dependent (related and more detailed computations can be found in \cite{fullcalc,monomial,Candelas:1975du,Miao:2006pn,Garbrecht:2006df,Parker:1999td}). Now we notice a potential subtlety:  if we are interested in a single field case, the inflaton has to be very light compared to the Hubble mass in order to generate density perturbations during inflation.  In this regime the argument of the logarithm will become negative for momentum modes outside the horizon. We have uncovered a problem with the standard importation of the Coleman-Weinberg calculation from flat space to the de-Sitter setting. This implies that for a light scalar field in de-Sitter, like the inflaton, the Coleman-Weinberg approach is not adequate. Instead some other approach must considered for dealing with the infrared loop effects during inflation \cite{Vilenkin:1982sg,Starobinsky:1994bd,Mukhanov:1996ak,Abramo:1998hi,Losic:2005vg,Sloth:2006az,Finelli:2001bn,Riotto:2008mv,Weinberg:2005vy,Urakawa:2009gb,Giddings:2010nc,Giddings:2010ui,Byrnes:2010yc,Seery:2009hs,Burgess:2009bs,Giddings:2011ze,Giddings:2011zd,Dai:2013kra}. Since the infrared modes of a light field like the inflaton only grows logarithmically (not exponentially like in the fast-roll case) and are suppressed by a factor $H^2/M_p^2$, their effect only becomes important on very long time scales of order $t\sim R_{ds} S_{ds}$, where $R\sim 1/H$ is the de-Sitter radius and $S_{ds} \sim (M_p/H)^2$ is the de-Sitter entropy \cite{Giddings:2010nc,Giddings:2011ze}. Furthermore, in single field inflation their effect have no local gauge-invariant implications and are only important for global questions like eternal inflation \cite{Giddings:2010nc,Giddings:2011ze}, or when looking for small statistical anisotropies by comparing different parts of the sky \cite{Giddings:2011zd,Dai:2013kra}. In any case, this problem is very orthogonal to the effect we wish to investigate, and we will not dwell on it here.  

For our purposes, we may reassemble all smaller wavelength modes into an effective potential with an IR cutoff at the horizon scale, and treat all perturbations as propagating in the shifted background. This procedure yields an effective action equivalent to (\ref{slaw}), so that we need not concern ourselves with this mass shift any more beyond this point, and perform all our loop calculations in flat space.  We are not permitted to do this same procedure for the case of fast-roll, where the effect comes in at wavelengths smaller than the horizon whose observable effects we are interested in calculating.  

\subsection{The complete scenario}\label{mediators}
We are finally ready to combine our results into a scenario that gives rise to large running to the inflaton $\phi$ through the explicit time dependence of a fast roll field $\chi$.  As explained above, $\chi$ cannot couple directly to the inflaton, or the exponential running would void the slow roll approximation within a few e-folds.  Hence, we introduce a third field $\rho$ that serves as a type of messenger between the two sectors.  As this field provides the setting for a large amount of running to take place, we refer to it as the {\it marathon}.  In order to embed the eigenvalue splitting Hessian discussed in section \ref{multi}, we want the marathon field to have the following properties:  (i) Its mass should be proportional to the fast roll field only and (ii) it must couple linearly to the inflaton, so that it will appear in the off-diagonal elements.  

We display again the general form of our potential (\ref{potentia}):
\beq
V(\phi,\chi,\rho)=V(\phi)-\frac{\mu^2}{2}\chi^2+f(\phi)\rho+g(\chi)\rho^2
\eeq
which leads to a Hessian
\begin{equation*}
V_{ij}=\begin{pmatrix}
V''+f''\rho & f' & 0\\
f' & 2g & 2g'\rho\\
0 & 2g'\rho & -\mu^2+g''\rho^2
\end{pmatrix}
\end{equation*}
Ideally, the VEV of the marathon field would vanish, so that the matrix would become block diagonal, isolating the negative eigenvalue that must be excised from the spectrum before computing the effective potential.  Thankfully, the field-dependent minimum of the marathon's potential occurs at 
\beq
\langle\rho\rangle=-\frac{f(\phi)}{2g(\chi)},
\eeq
 a function of the fields\footnote{The reader may be puzzled why we don't just complete the square and have a funny potential for the inflaton and fast roll field, uncoupled from the marathon.  If we did this the standard formula we have written for the Coleman-Weinberg potential would be invalid, and must be replaced with the field redefinition invariant expression introduced by Vilkovisky \cite{vilko}.  In this parameterization the kinetic terms become nonminimal, though we can restrict to the regime where they are approximately canonical, as in \cite{mutated}. The ``connection terms" enforce that the effective potential is the same for both cases, but the treatment is ultimately simpler as it is originally written.}.  In the first case we consider $g$ is a positive power of $\chi$, and so the VEV is dynamically driven to zero.  In section \ref{neginv} we consider negative powers of $\chi$, which leads to a large VEV for the marathon, but the combination $g'\langle\rho\rangle$ that appears in the Hessian is still small.  Otherwise, the eigenvalues of this $3\times3$ matrix must be computed, but the results for this case are not shown explicitly, as they totally obfuscate the physical picture.

In general, there are a list of constraints this matrix must satisfy in order for the loop effects to resemble a pure running in the inflaton sector.  There are a number of potential hazards that could render other effects more important than the running of the spectral index, which we must ensure are negligible on a case by case basis.  We enumerate the general pitfalls in this section, and display which are the most dangerous in each of the specialized scenarios.

(i) Slow-roll will be satisfied as long as
\beq
f''\rho\ll V''
\eeq
otherwise the inflaton would pick up a large mass from its interactions with the marathon, even at tree level, and inflation would not occur.

(ii) The off-diagonal components will be perturbatively small if
\beq
f'\ll g,\quad g'\rho\ll g
\eeq
The first of these is required for the level-splitting mechanism as in (\ref{splitting}) to occur.  The second is for convenience, so that the fast roll eigenvalue of the Hessian can be easily isolated.

(iii) There is a stronger condition on the off diagonal components that ensures a runaway direction does not develop by the mixing of the inflaton and marathon:
\beq
f'^2\ll V''g
\eeq
This is likely to be violated before condition (ii).  In this case there is maximal mixing in the inflaton-marathon sector, and only a single positive eigenvalue.  This does not necessarily preclude a running spectrum, as it only affects whether the subdominant eigenvalues can be interpreted as an effective potential, but we choose to stay away from this regime for simplicity.

(iv) Our mechanism requires fast roll to be maintained for sufficiently long.  This will not be spoiled if 
\beq
g''\rho^2\ll \mu^2,\quad g^2\ll\mu^2\chi^2
\eeq
Where the first condition ensures that the contribution to the $\chi$ field's mass that comes from interacting with the marathon is subdominant.  The second condition ensures that one loop effects do not dominate the tree level terms in the $\chi$ sector.

(v) We also must require that the marathon field is heavy but sub-planckian,
\beq
H^2\ll g\ll M_p^2
\eeq
otherwise the marathon will not track its minimum, and will behave as an additional light field, contributing isocurvature modes to the power spectrum.

(vi) In addition, the energy density in the fast roll field must remain negligible,
\beq
\mu^2\chi^2\ll V_0
\eeq
Which will put an upper bound on the number of e-folds fast roll that can be sustained.  For $\mu\approx H$, this condition always reduces to 
\beq
F\left(\frac{\mu^2}{H^2}\right)N_{\text{TOT}}<\log\frac{M_p}{H}
\eeq
where $N_{\text{TOT}}$ is the total number of e-folds of fast roll, and $F\sim\mu/H$ is defined in (\ref{defF}).  Notice that if we take the quartic regulator of the fast roll field to be purely induced by quantum gravitational effects, precisely the same condition is yielded.  The absolute minimum we can take the Hubble rate during inflation to be is set by demanding that the energy density $M_p^2H^2>(10\, {\rm TeV})^4$, corresponding to $FN_{\text{TOT}}<66$.

If all of these conditions are satisfied, then the system behaves as a 1-loop potential for the inflaton:
\beq\label{pot1}
V_{\text{1-loop}}=\frac{1}{64\pi^2}\left[(m_+^2)^2\left(\log(m_+^2)-\frac32\right)+(m_-^2)^2\left(\log(m_-^2)-\frac32\right)\right]
\eeq
Where $m_\pm^2$ are the eigenvalues of the $2\times2$ Hessian.  The general expression for this was given in (\ref{eigem}).  The eigenvalues become
\beq
m_+^2\approx 2g+\frac{f'^2}{2g},\quad m_-^2\approx V''-\frac{f'^2}{2g}
\eeq
but since $m_- << m_+$, the contribution to the potential (\ref{pot1}) proportional to $(m_-^2)^2$ is suppressed, and we can therefore neglect it. In the $m_+^2$ term, however, the $\chi$ term will dominate the logarithm, and in the cross term the $\chi$ dependence will completely cancel!  The subleading dependence in the logarithms will slightly alter the constant offsets, so that the final expression for the effective potential is

\beq
V_{\text{1-loop}}=\frac{1}{64\pi^2}\Bigg[4g^2\left(\log2g-\frac32\right)+2f'^2\bigg(\log2g-1\bigg)\Bigg]+\mathcal{O}\left(\frac{f'^4}{g^2}\right)
\eeq

The first term only involves the $\chi$ field, and as long as it does not destroy the fast roll behavior it does not contribute to observables.  The second term provides the proxy running in the power spectrum.  Since the masses in this equation are much larger than the Hubble rate, we can safely neglect curvature corrections at this point.

\subsection{Simplest example: Positive running}\label{simple}
Now we specialize to the potential of section \ref{mediators}, and the mixing functions to $f(\phi)=\Lambda\phi^2$ and $g(\chi)=n\chi$, surely an innocuous choice of renormalizable operators.  The Hessian becomes

\begin{equation*}
V_{ij}=\begin{pmatrix}
m^2+2\Lambda\rho & 2\Lambda\phi & 0\\
2\Lambda\phi & 2n\chi & 2n\rho\\
0 & 2n\rho & -\mu^2
\end{pmatrix}
\end{equation*}
As stated before, if one of the fields is in the fast roll regime, then one of the eigenvalues of this matrix will be negative, and it should be excluded from the computation of the Coleman-Weinberg effective potential, in favor of using the Weinberg-Wu correlator.  In the limit that $\rho\rightarrow 0$, this excision is trivial because the matrix becomes block diagonal.  This is quite fortuitous for us, because we can see that the $\rho$ field actually has a minimum of its potential $\rho_{min}=-\Lambda\phi^2/(2n\chi)$, and since the $\chi$ field is exponentially growing, this is driven to be very small dynamically.  We will discuss the validity of this approximation in more depth below, but for now assume the negative mode decouples and focus on the upper $2\times2$ block.

The result is
\beq
V_{\text{1-loop}}\approx\frac{\Lambda^2}{8\pi^2}\phi^2\bigg(\log(2n\chi)-1\bigg)\label{direct}
\eeq
So that 
\beq
\Delta m^2=\frac{\Lambda^2}{4\pi^2}F\left(\frac{\mu}{H}\right)\label{expressyoself}
\eeq

Notice the unavoidable prediction from this potential that the running is positive.  Below we give two modifications to this scenario that give negative running.

Let us list the most stringent constraints on this scenario from section (\ref{mediators}).  We take the value of the fast roll field to be $\chi=\mu e^{F\Delta N}$ at CMB scales, so that the VEV can be large enough for the mass of the marathon to dominate.  Then, using the fact that $\phi\approx HN/\sqrt{P_k}$ (\ref{prophile}) and $\Lambda\sim\sqrt{4\pi^2\eta}H$, find numerical conditions for the remaining parameters.  The dominant constraints from the previous section are (iii), which is the absence of inflaton-marathon mixing, and (v), keeping the marathon mass sub-planckian. Together, these yield,
\beq
25<K+F\Delta N<2\log\frac{M_p}{H}
\eeq
with $K=\log n/H$.  This has to be negative to not give a large positive contribution to the fast roll field.
Consistency here demands that $H\lesssim10^{-6}M_p$.  The energy condition (vi) implies
\beq
F N_{\text{TOT}}<\log\frac{M_p}{H}
\eeq
This is quite a stringent bound, given that the minimum possible value for $N_{\text{TOT}}=\Delta N+8$, corresponding to fast roll ending immediately after CMB scales.  Nevertheless, if $F\sim 1$ and $K\sim-3$, we arrive at $H<10^{-17}M_p\sim10 \text{GeV}$, corresponding to an energy density $\rho\sim(10^9\, \text{GeV})^4$, a low but acceptable scale of inflation.  The maximum we can allow for $F$, based on the consideration that the energy density during inflation $M_p^2H^2>(10\,\text{TeV})^4$, would be $F_{max}\approx 5.5$, corresponding to $\mu\approx7H$.

Our figure of merit
\beq
\frac{\alpha_s}{n_s-1}=\frac{2F}{K+F(\Delta N-\Gamma_0)+3.6-\frac{4\pi^2 m^2}{\Lambda^2}}
\eeq
interpolates between $0$ and $2/\Delta N$, if we insist that $n_s<1$.

\subsection{Negative running from inverse powers of fields}\label{neginv}
One method of obtaining negative running is to consider potentials that are nonanalytic in the field $\chi$.  These types of potentials were first considered in \cite{rolling}, and appear naturally in brane inflation \cite{brane}, where they represent higher dimensional analogues of the Coulomb potential.  Otherwise, it is possible to arrive at these potentials from integrating out a heavy field, as seen in \cite{mutated}.  A candidate potential is
\beq
V(\phi,\chi)=V(\phi)-\frac{\mu^2}{2}\chi^2+\Lambda\phi^2\rho+\frac{\kappa^{2+q}\rho^2}{\chi^q}
\eeq
Which corresponds to the choice $f(\phi)=\Lambda\phi^2$ and $g(\chi)=\kappa^{2+q}/\chi^q$.   The logic behind this choice is that the Hessian will be more or less the same as in the case above, except that now the $V_{\rho\rho}$ term will depend on an inverse power of the field $\chi$, which will provide a minus sign once the logarithm is taken.  The Hessian is
\begin{equation*}
V_{ij}=\begin{pmatrix}
V''+\Lambda\rho & 2\Lambda\phi & 0\\[1mm]
2\Lambda\phi & 2\frac{\kappa^{2+q}}{\chi^q} & -2q\frac{\kappa^{2+q}\rho}{\chi^{1+q}}\\[2mm]
0 &  -2q\frac{\kappa^{2+q}\rho}{\chi^{1+q}} & -\mu^2+q(q+1)\frac{\kappa^{2+q}\rho^2}{\chi^{2+q}}
\end{pmatrix}
\end{equation*}
If we consider the one loop potential for this scenario, then indeed we get a term in the potential
\beq
V_{\text{1-loop}}=-\frac{1}{8\pi^2}\Lambda^2\phi^2\left(\log\left(\frac{\chi^q}{2\kappa^{2+q}}\right)+1\right)
\eeq
with the minus sign that gives negative running.  It is still necessary to make sure that the tree level terms do not interfere with any of the dynamics in an appreciable way, but this is easy to accomplish if we take the mass scale $\kappa$ to be sufficiently large.  The VEV of the marathon field now grows as $\langle\rho\rangle=-\Lambda\phi^2\chi^q/2\kappa^{2+q}$, but this is fine, as the potentially complicating mixing between the fast roll field is suppressed by an additional three powers of $\chi$ and so is driven to 0.  

We now study the constraints: the position of the fast roll field must be far enough down its potential that the inverse powers do not dominate from condition (iv), yet it cannot be too far down, or else there will be too much mixing between the inflaton and marathon (i), (iii).  These give the following bound:
\beq
\frac{2+q}{1+q}L< F\Delta N<\frac{2+q}{q}L-\frac{24}{q}
\eeq
where $L=\log(\kappa/H)$.  This condition has the possibility of becoming violated near the end of the fast roll phase, in which case there will be additional effects in the power spectrum, but we enforce that this effect is negligible in this work.  Consistency of these conditions enforce that $L>24(1+q)/(2+q)$, and if we want mixing effects to be absent in the CMB we need to have that not only the initial value of $\chi$ be within this window, but also the value at the scale $\ell\approx2500$, corresponding to $\Delta N+8$.  We can see that the value of $q$ does not matter much, and indeed in the lower bound on $F\Delta N$ its dependence cancels completely.  For lower values of $q$, however, it is possible to make $L$ smaller and still satisfy the constraints over the entire range of CMB scales.  Then for definiteness we take $q=1$ and $L=24$.  In this regime the other constraints are trivially satisfied, except for (vi):
\beq
36+8F<\log\frac{M_p}{H}
\eeq
If we take $F=1$ and $\Delta N=36$, then $H\sim \text{GeV}$, a low but technically acceptable scale of inflation.  This can be relaxed if other effects in the power spectrum are tolerated, but will require a more detailed calculation.
Our figure of merit is now
\beq
\frac{\alpha_s}{n_s-1}=\frac{-2qF}{(2+q)L+qF(\Delta N-\Gamma_0)+3.6-\frac{4\pi^2m^2}{\Lambda^2}}
\eeq
For the values we have chosen, this is equal to $1/10$ when  $m^2\approx 3.3\Lambda^2$.

\subsection{Negative running from fermions}\label{negferm}
We describe an alternative, perhaps simpler, method of obtaining negative running.  It is quite natural to get the correct sign by integrating out heavy fermions in the loops instead \cite{cole,susyinf,Buchmuller:2014epa}.  In this case we require two fermions (marathinos?) to act as messengers, and want to reproduce the Hessian matrix we had before.  To do this, we note that if the potential is of the form
\beq
V(\phi,\chi,\psi_i)=V(\phi)+V(\chi)+m_{ij}(\phi,\chi)\bar{\psi}_i\psi_j+\text{h.c.}
\eeq
Then the corresponding 1-loop potential will be 
\beq
V_{\text{1-loop}}=-\frac{1}{64\pi^2}\text{Tr}(m^\dagger m)^2\log(m^\dagger m)
\eeq
This seems to present a problem at first glance, as now there is not as much freedom in choosing the elements of the matrix $m^\dagger m$ as there was in the scalar case, since it is the square of an unconstrained matrix.  In the simplest case of a model with two fermions, and taking the couplings to all be real for simplicity, the matrix becomes
\begin{equation*}
m^\dagger m=\begin{pmatrix}
m_{11}^2+m_{12}^2 & m_{12}(m_{11}+m_{22}) \\
m_{12}(m_{11}+m_{22}) & m_{12}^2+m_{22}^2
\end{pmatrix}
\end{equation*}
To be in the situation we were before, we need the off diagonal elements to depend linearly on $\phi$, and the dominant diagonal element to depend on $\chi$.  However, all of the matrix elements appear in the off-diagonal terms, and so if the fast roll field dominates the diagonal elements it will also appear multiplying $\phi$, leading to an exponentially fast running.  

However, there is a rather trivial workaround of this issue, if one of the fields couples to the fermions along with the matrix $\gamma^5$.  If $m=m_1+i\gamma^5m_2$, then $m^\dagger m=m_1^\dagger m_1+m_2^\dagger m_2$, and the matrix that appears in the one loop potential becomes a sum of two terms.  For example, we can take the tree level potential
\beq
V(\phi,\chi,\psi_i)=V(\phi)+V(\chi)+\alpha\chi\bar{\psi}_2\psi_2+i\beta\phi\bar{\psi}_2\gamma^5\psi_2+i\Lambda\bar{\psi}_1\gamma^5\psi_2+\text{h.c.}
\eeq
In this case the Hessian will be
\begin{equation*}
m^\dagger m=\begin{pmatrix}
\Lambda^2 & \beta\Lambda\phi \\
\beta\Lambda\phi & \alpha^2\chi^2+\Lambda^2+\beta^2\phi^2
\end{pmatrix}
\end{equation*}
which has manifestly positive eigenvalues, and the previous mechanism can be invoked.

Note that the 1-loop potential also induces a $\chi^4$ term, which will spoil fast roll if 
\beq
\alpha^{2}\ll e^{F(\mu/H)N_{\text{TOT}}}
\eeq
as this is proportional to the square of the term in the Lagrangian, it is generically less stringent than the bare term $\lambda\chi^4$ we should have included.  Still, this tells us that the marathon field has to be weakly coupled to the fast roll field for this mechanism to work.  In general, though, this scenario has less constraints to satisfy, mostly due to the fermionic nature of the additional fields preventing any substantial mixing.  The constraints (i)-(v) give
\beq
\text{max}\bigg\{0,10+\log\left(\frac{\beta\Lambda}{H}\right)\bigg\}<\log\alpha+F\Delta N<\log\frac{M_p}{H}
\eeq
Large running will occur when $\Delta m\sim m$, corresponding to $\beta\sim 1$.  For definiteness we take $\beta=1/e$, $F=1$, and $\Delta N=26$, then these constraints can be satisfied, with $\alpha\sim10^{-7.4}$ and $H<10^{-15}M_p\sim \text{TeV}$.

\subsection{Curvaton induced running}\label{curvaton}
Let us consider the case where the running of the spectral index is induced by the intrinsic running of a curvaton \cite{curv}. In this case the curvaton contributes some fraction, $f$, to the final curvature perturbation spectrum \cite{Langlois:2004nn,Kinney:2012ik,Fonseca:2012cj,Enqvist:2013paa,Byrnes:2014xua}, which can be parametrized as 
\beq\label{specttotal}
P(k) = A_s\left[(1-f)\left(\frac{k}{k_*}\right)^{(n_{inf}-1)}+f\left(\frac{k}{k_*}\right)^{(n_{\sigma}-1)}\right]~,
\eeq
and the intrinsic running of the curvaton is given by
\beq
\alpha_{\sigma} = \frac{d n_{\sigma}(k)}{d \ln k}~.
\eeq
Assuming that $\alpha_{\sigma} \gg (n_{inf}-1)^2 \sim (n_{\sigma}-1)^2$, the running of the spectrum (\ref{specttotal})  is simply
\beq
\alpha_s \approx f \alpha_{\sigma}~.
\eeq
Similarly the tensor-to-scalar ratio will given by
\beq
r = (1-f) 16 \epsilon~.
\eeq

We will assume that the intrinsic running of the curvaton is of the ``running by proxy" form\footnote{See \cite{Takahashi:2013tj,Peloso:2014oza,curv} for other proposals of curvatons with intrinsic running.}, and the curvaton potential has a logarithmic correction of the form
\beq\label{cV1}
V_{\text{1-loop}}(\sigma) = -\frac{1}{2}\Delta m_\sigma^2\log(\chi/\chi_0)\sigma^2
\eeq
where the potential of the proxy field is
\beq
V(\chi) = V_0 - \frac{1}{2} m_\chi^2 \chi^2~.
\eeq
The logarithmic correction to the curvaton potential form the proxy field in (\ref{cV1}) follows from a logic similar to the one which led to the proxy term in the inflaton potential, except that the general shape of the potential is less constrained in this case. In particular, let us consider the constraint $V_N/V \ll V_N''/V$ which led us to the conclusion that inflation must be of the small field type in order to be consistent with proxy running (see section \ref{proxrun}). The equivalent of this constraint in the curvaton case becomes $V'_N/V \ll (\p_\sigma^2 V_N)/V$, which is naturally satisfied  even when the inflation is of the large field type $\phi\gg M_p$, since $\sigma \ll M_p$ unless we are dealing with an inflating curvaton \cite{Langlois:2004nn}. Therefore, large field inflation is also naturally compatible with curvaton induced running of the proxy form. If the universe f.ex. originated from chaotic inflation with a simple quadratic potential $V(\phi) = \frac{1}{2} m^2\phi^2$, then a fifty percent curvaton fraction, $f= 0.5$ with an intrinsic running of $\alpha_{\sigma} =-0.1$, would lead to a final curvature spectrum with $\alpha_s = -0.05$ and $r=0.07$. 

In the case of a future simultaneous detection of running and large primordial tensor modes, one is surprisingly lead to the conclusion that it is only consistent with proxy running if it is induced by a curvaton in a large field model. If a large anomalous running is detected, but tensor modes are constrained to be small, we are are lead to the conclusion that this is both consistent with proxy running of the inflaton or running induced by proxy running of the curvaton. One possible way of discriminating between these two cases could be through an enhanced non-Gaussian component of the curvature perturbation (consistent with current observational bounds) in the curvaton case \cite{curv,Enqvist:2014nsa}. 

\subsection{Super-horizon evolution of perturbations} \label{superh}

A potentially important feature of this model is the small mixing between the inflaton and the fast-roll field that is generically induced. When the fast-roll regime has ended and the fast-roll field has decayed, the final total curvature perturbation will simply be given by the inflaton field perturbation  
\beq
\zeta = \zeta_\phi =\frac{H}{\dot \phi}\delta\phi~. 
\eeq
One might however worry that the exponentially growing perturbations of the fast-roll field will source perturbations in the inflaton field before the decay of the fast-roll field due to the mixing between the two fields. If we consider the general equation of motion for the field perturbations in the spatially flat gauge (letting the roman index denote the different field species, i.e.  inflaton, marathon or proxy field)
\beq
\ddot{\delta\phi_i}+3H\dot{\delta\phi}_i+\frac{k^2}{a^2}\delta\phi_i+\left[V_{ij}-\frac{1}{M_p^2}\frac{1}{a^3}\frac{d}{dt}\left(\frac{a^3}{H}\dot\phi_i\dot\phi_j\right)\right]\delta\phi_j=0 ~,
\eeq
we find a gravitational mixing between $\delta\phi$ and $\delta\chi$ from the second term in the square-bracket, and a direct mixing from the first term in the square-bracket. We want to require that the effect of the exponentially growing perturbations of the fast-rolling proxy field, $\delta\chi$, on the inflaton perturbations, $\delta\phi$ remains small. Following an analysis similar to that of \cite{Ferreira:2014zia}, if the gravitational mixing dominates we find the constraint
\beq
F (N_{\text{TOT}} +N) < -10 +2\log\frac{M_p}{H}
\eeq
and if the direct interaction  in \ref{direct} dominates, we find 
\beq
F\Delta N > -1 +\log \frac{\phi}{H}+\log\frac{\Delta m^2}{H^2}+\log\frac{N_{\text{TOT}}}{F}~.
\eeq
Using $\phi\sim N H/\sqrt{P_k}$ both constraints are weaker than those discussed in the sections above.

\subsection{Stability to higher loops}\label{higher}

We make a few brief comments about the few approximations we took in deriving the one loop Coleman-Weinberg effective potential.  Firstly, we note that to the accuracy at which we calculate, it is completely justified to truncate the effective potential to one loop.  In general the two loop contribution will contain terms of the same form as the ones we arrived at $V_2\sim m^4\log(m^2)$, but the entire correction is suppressed by an additional loop factor of $1/(16\pi^2)$ \cite{2loop}.  This is smaller than the slow roll parameters, to which we have worked at first order, so that at this stage it actually would have been inconsistent to endeavor a two loop calculation.  Note as well that $\log(\chi)^2$ terms will appear in the two loop potential, leading to a running of the running a factor of $\mathcal{O}(\frac{1}{100})$ smaller than the running itself.  This is still larger than the usual single field prediction that the running of the running be third order in slow roll parameters, but nevertheless unobservably small.

A second approximation we made was to neglect graviton loops in our effective potential.  These will lead to two distinct types of contributions \cite{fullcalc}: terms that schematically appear as $R^2$, and terms that appear as $\log(R)$.  The higher curvature terms must be neglected in the background evolution, as they introduce apparent spurious degrees of freedom not present in the low energy theory, which must be removed by hand.  The same goes for the logarithms, which must be expanded in a Taylor series and truncated at first order, which effectively yield corrections to the Planck mass.  These corrections will be of the order $m^2/M_p^2$, $\phi_i^2/M_p^2$, and so will be completely negligible for our (small field) purposes.

Finally, we comment on the radiative stability of our choice of potential in (\ref{potentia}).  The absence of the fast roll field in the linear marathon sector may be a bit disconcerting, especially since this was crucial for our mechanism to be operational.  We provide no underlying motivation for why this arrangement may have come about from a particle physics viewpoint, but we do note that this scenario is internally consistent, in that higher order loops do not induce this dangerous term. 
\begin{figure*}[thb]
\centering
\includegraphics[width=5cm,height=2.8cm]{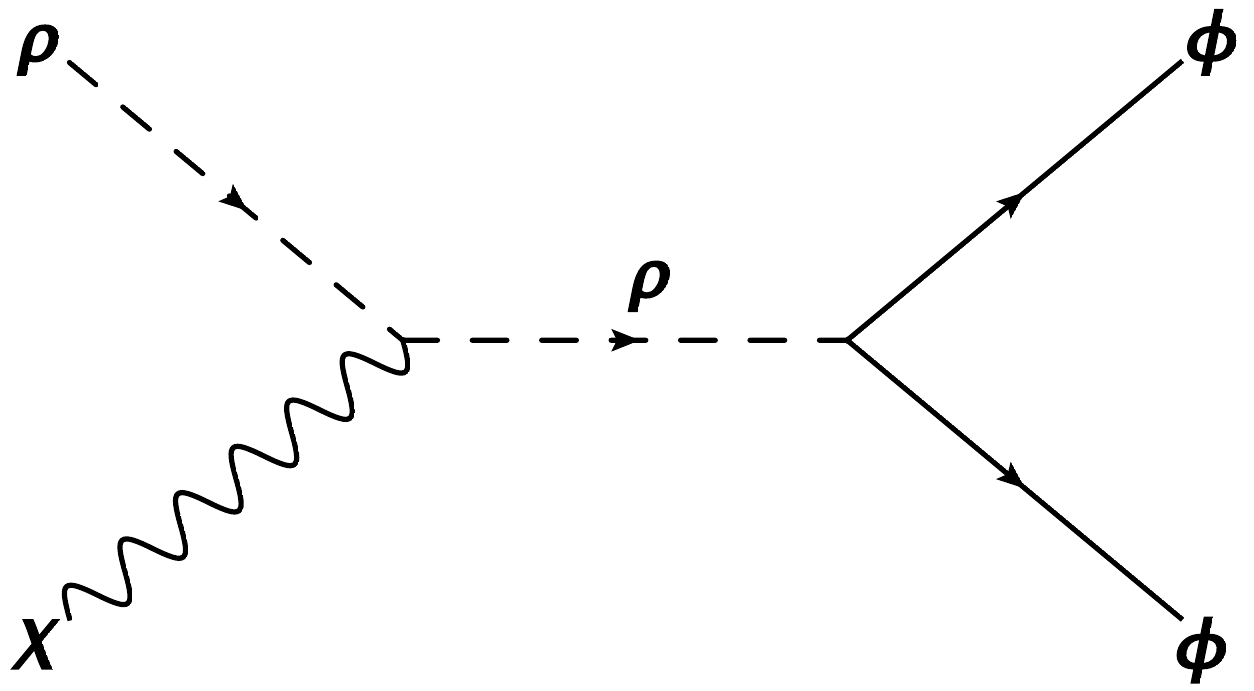}
\hskip 10pt
\includegraphics[height=2.5cm]{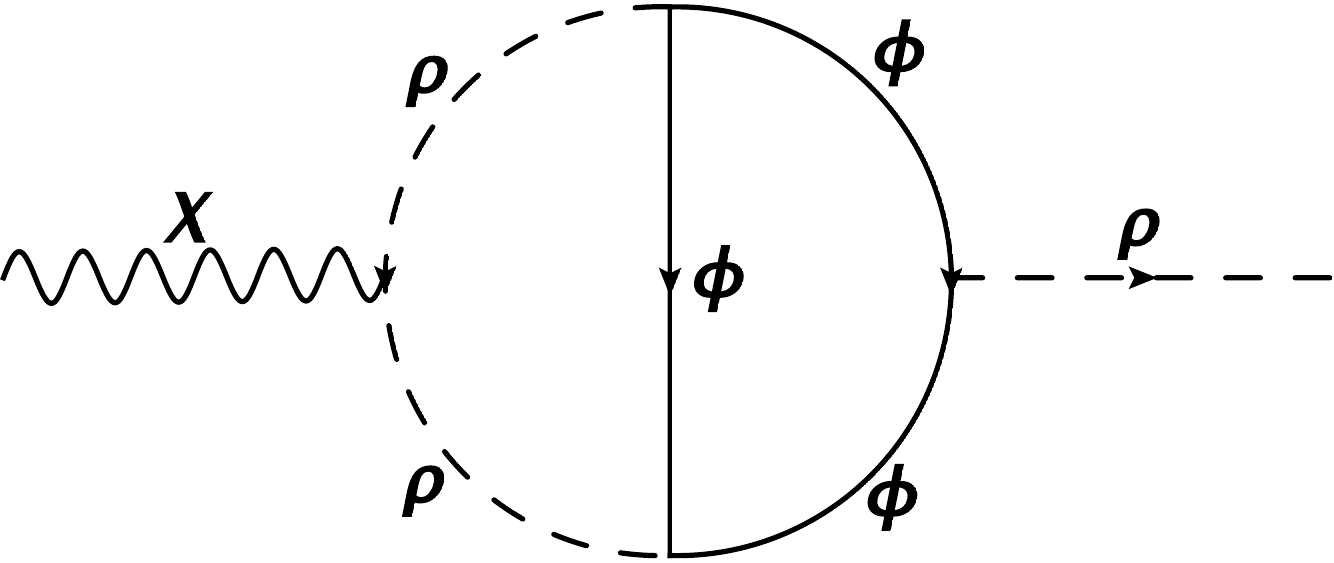}
\hskip 10pt 
\includegraphics[height=3.2cm]{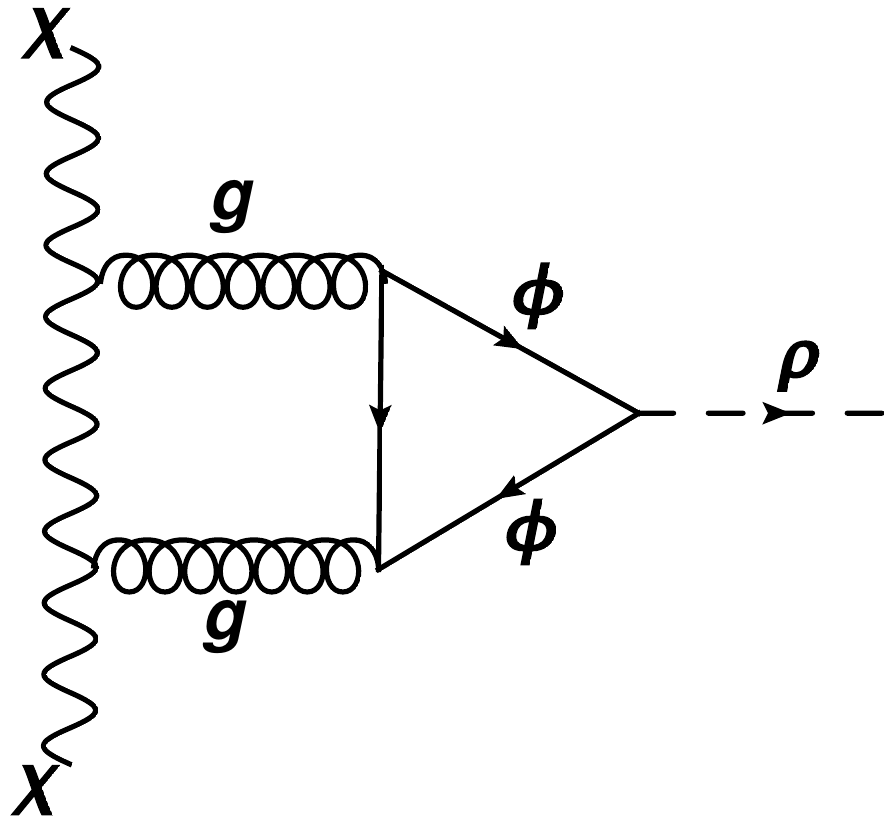}
\caption{Various loop contributions of the form $V\propto\chi^n\rho$}
\label{diagrams}
\end{figure*}
To begin with, we note that in the potential there are two vertices, one with two $\phi$s and one $\rho$ proportional to $\Lambda$, and one with one $\chi$ and two $\rho$s proportional to $n$, as well as gravitational interactions.  We would like to estimate the leading contribution that will have external $\chi$ legs and a single external $\rho$.  The first worrisome diagram is depicted in \ref{diagrams}(a), and is a tree level diagram in which an internal marathon field is integrated out to yield a contact interaction term $V_{\rm eff}\sim n\Lambda\chi\phi^2\rho/m_\rho^2$, which at first glance seems to yield dangerous fast roll-marathon coupling.  However, notice the presence of the marathon mass in the denominator, the dominant contribution to which is $m_\rho^2\sim n\chi$.  Thus, the $\chi$ dependence completely cancels out, and we are left with a term of the form we had to begin with.  Let us briefly acknowledge that the situation is slightly more subtle than we have presented:  the fact that the mass of $\rho$ is a time dependent quantity that depends on other fields means that we are not actually allowed to use the methodology of Lorentz invariant field theory, but must instead calculate corrections to the quantum fluctuations of perturbations on top of the background field values.  These corrections for the fluctuations will in general be different from the corrections to the equations the background fields obey, which will cause the resulting effective Lagrangian to not be a simple function $L(\bar\phi+\delta\phi)$, but instead a more general $L(\bar\phi,\delta\phi)$.  Nevertheless, we have approximated this full form as the former by integrating out the marathon field in a flat space setting, and then naively applying those results to the case at hand.  This approximation will only maintain its validity inasmuch as the fluctuations remain tightly coupled with the background fields, but, given the tiny coefficients multiplying the corrections we find, we take this simplification to be an adequate estimate for the time being.

We see that to estimate the form of the effective potential we must count powers of $\phi$, $\rho$, and $\Lambda$, which cannot be altered by inverse propagators, and then enforce that the total dimension of the term in the potential be 4.  If there are other internal lines the dimensions can be balanced with other propagators that do not remove any $\chi$s, but note that there is always at least one $\rho$ propagator (excluding gravitons), and so at least one potential cancellation.  Dangerous diagrams will have less powers of external inflatons, which necessarily entails they enter exclusively through loops, which is also beneficial for the denominator to contain purely constant dimensionful quantities.  The simplest diagram would be to close the inflaton legs in \ref{diagrams}(a), but this is taken to be zero by the no tadpole condition (another importation from flat space we employ).  Instead, the simplest diagram is depicted in \ref{diagrams}(b), which is a two loop contribution.  The full expression will not be written down, but we can estimate the diagram as $V_{\text{eff}}\sim n\Lambda^3\int d^{4}k/(k^2)^5 \chi\rho$.  The full integral is a sum of several terms, for most of which the $\chi$ dependence drops out completely, or is at most logarithmic.  The one dangerous contribution comes from the region of the integral where the momentum of the internal inflaton dominates, proportional to $1/m_\phi^2$.   The estimation for the effective potential, including loop factors, is then
\beq
V_{\text{2-loop}}\approx\frac{1}{(16\pi^2)^2}\frac{n\Lambda^3}{m^2}\chi\rho
\eeq
This leads to a mass mixing of the fast roll field and the marathon that cold potentially rival the $\rho$ dependent mixing present in the tree level potential.  The effect of this induced term can be estimated by comparing it to $n\chi\rho^2$
\beq
\frac{V_{\text{2-loop}}}{V_{\text{tree}}}\approx\frac{1}{(16\pi^2)^2}\frac{\Lambda^3}{m^2\rho}\approx 10^{-4}\frac{n\Lambda^2\chi}{m^2\phi^2}
\eeq
In terms of variables defined in section \ref{neginv}, this translates into
\beq
K+F\Delta N<30.
\eeq
Which is satisfied for our choice of parameters $K+F\Delta N\sim 25$, in which case the importace of this term is suppressed, essentially by the loop factors.  This term does not contribute significantly to the dynamics of the theory.

There is another diagram that appears at 2 loops if we consider internal graviton propagators, as in \ref{diagrams}(c).  In this the coupling of the graviton to the fast roll field with vanishing external momentum is $\sim\mu^2/M_p$, while the most relevant coupling to the virtual inflaton comes with two powers of momentum, to completely balance the total six propagators.  Roughly, the diagram evaluates to 
\beq
V_{\text{eff}}\approx \frac{1}{(16\pi^2)^2}\frac{\mu^4}{M_p^4}\Lambda\chi^2\rho\lesssim\left(\frac{H}{10M_p}\right)^4\Lambda\chi^2\rho
\eeq
With the inequality saturated for the models that have the highest possible inflation scale, and moderately tolerable tuning to get enough fast roll.  This is even smaller than the contribution from scalar loops.  These loops yield the dominant effects, and so our scenario is stable to quantum corrections.

\section{Conclusions}\label{concl}

We are currently at a very interesting moment of inflationary cosmology. Our experimental knowledge has gone from allowing almost every conceivable model of inflation to beginning to rule out some of the simplest models of inflation, but yet data does not have the precision to pinpoint one specific model as the unambiguously preferred. This makes it particularly important to think about which observables can be used to discriminate between different models. The tensor-to-scalar ratio and non-Gaussianity are important observables, and have both received much attention in the past for this reason. On the other hand the running of the spectral index has received less attention, since there are very few existing models which are capable of generating a large running compared with the general expectation in (\ref{runninggen}), and so far they have all been large field models\footnote{With the exception of the curvaton models \cite{Takahashi:2013tj,falsefat,curv}}. However, the tensor-to-scalar ratio already appears as a good observable for testing large field models, and in any case the current constraints on running are not far from the expected value in this class. 

On the other hand, small field models with a vanishing tensor-to-scaler ratio and an observable level of running are truly exotic from the point of view of the general expectation in (\ref{runninggen}). Therefore, in the future we might well be in an experimental situation where we see a vanishing tensor-to-scalar ratio, no evidence of non-Gaussianity, and hence have no expectation of observing running of the spectral index. However, we do not believe that such a situation will deter experimentalists from continuing to improve the constraints on the running, being one of the limited number of cosmological observables, but it may be reasonable in this case to ask which kind of new physics could lead to an observable running of the spectral index in an effective single field model of small field inflation. 

We have presented a simple class of models capable of giving a large amount of running of the spectral index of the inflaton in the small field regime by using the general mechanism of proxy running. In the proxy model, the running can be viewed as coming from an explicitly time-dependent single field potential, but at the microscopic level the time-dependence is derived from integrating out heavy fields. To accomplish this we invoke at least two additional fields, one of which is in fast roll, and another that serves only as a messenger for loop effects.  For our choice of potentials, the Coleman-Weinberg loop resummation then induces an explicit time dependence in the parameters of the inflaton potential that is exactly linear in number of e-folds, leading to an observable amount of running that can be up to the same value as the spectral tilt.  

Another interesting outcome of our analysis is that proxy running in an effective single field model is only compatible with a small field models of inflation. Therefore, any future observation of a non-vanishing tensor-to-scalar ratio will imply that an anomalously large running can only be compatible with proxy running, if the proxy running is in a curvaton sector that contribute only some fraction to the final spectrum and induces running this way  \cite{curv} . This conclusion is a bit surprising, since curvaton models usually are thought to be associated with a vanishing tensor-to-scalar ratio. One possible feature of this scenario is an interesting signature of non-Gaussianity  \cite{curv,Enqvist:2014nsa} . 

As a period of fast roll is quite difficult to maintain for a significant number of e-folds, we expect that our model will exhibit features associated with the end of fast roll at some point very near the current observable window.  Depending on the precise parameters we choose, there may also be a significant amount of mixing between the inflaton and fast roll correlators, with definite consequences for the power spectrum.  There are many generalizations of this simple model that can be made, such as more specialized types of couplings, and considering implications for other types of correlators.  It would also be of some interest to find an embedding of this model into a more UV complete setup.  We plan to return to these questions in the future.

%
%
%


${}$\\
{\bf \noindent Acknowledgements}

\smallskip

We thank Alessandro Codello for some general discussions. RKJ acknowledges financial support from the Danish council for independent research in Natural Sciences for part of the work.
We would like to thank the Lundbeck foundation for financial support.


\begin{thebibliography}{99}

\bibitem{inflation}
  A.~H.~Guth,
  Phys.\ Rev.\ D {\bf 23}, 347 (1981);
  A.~D.~Linde,
  Phys.\ Lett.\ B {\bf 108}, 389 (1982);
  B {\bf 114}, 431 (1982);
  B {\bf 116}, 335 (1982);
 \ B {\bf 116}, 340 (1982);
  A.~Albrecht and P.~J.~Steinhardt,
  Phys.\ Rev.\ Lett.\  {\bf 48}, 1220 (1982).
\bibitem{Starobinsky:1979ty} 
  A.~A.~Starobinsky,
  JETP Lett.\  {\bf 30}, 682 (1979)
  [Pisma Zh.\ Eksp.\ Teor.\ Fiz.\  {\bf 30}, 719 (1979)].

   \bibitem{planckxvi} 
  P.~A.~R.~Ade {\it et al.}  [Planck Collaboration],
  Astron.\ Astrophys.\  {\bf 571}, A16 (2014)
  [arXiv:1303.5076 [astro-ph.CO]].
  
\bibitem{Hinshaw:2012aka} 
  G.~Hinshaw {\it et al.}  [WMAP Collaboration],
  Astrophys.\ J.\ Suppl.\  {\bf 208}, 19 (2013)
  [arXiv:1212.5226 [astro-ph.CO]].
  
  \bibitem{spt} 
  Z.~Hou, C.~L.~Reichardt, K.~T.~Story, B.~Follin, R.~Keisler, K.~A.~Aird, B.~A.~Benson and L.~E.~Bleem {\it et al.},
  Astrophys.\ J.\  {\bf 782}, no. 2, 74 (2014)
  [arXiv:1212.6267 [astro-ph.CO]].
  
\bibitem{Sievers:2013ica} 
  J.~L.~Sievers {\it et al.}  [Atacama Cosmology Telescope Collaboration],
  JCAP {\bf 1310}, 060 (2013)
  [arXiv:1301.0824 [astro-ph.CO]].
  
\bibitem{Abell:2009aa} 
  P.~A.~Abell {\it et al.}  [LSST Science and LSST Project Collaborations],
  arXiv:0912.0201 [astro-ph.IM].
  
\bibitem{Laureijs:2011gra} 
  R.~Laureijs {\it et al.}  [EUCLID Collaboration],
  arXiv:1110.3193 [astro-ph.CO].
  
  


\bibitem{Enqvist:2001zp}
  K.~Enqvist and M.~S.~Sloth,
  Nucl.\ Phys.\ B {\bf 626} (2002) 395
  [hep-ph/0109214].

\bibitem{Lyth:2001nq} 
  D.~H.~Lyth and D.~Wands,
  Phys.\ Lett.\ B {\bf 524}, 5 (2002)
  [hep-ph/0110002].

\bibitem{Moroi:2001ct} 
  T.~Moroi and T.~Takahashi,
  Phys.\ Lett.\ B {\bf 522}, 215 (2001)
  [Erratum-ibid.\ B {\bf 539}, 303 (2002)]
  [hep-ph/0110096].
\bibitem{Mollerach:1989hu} 
  S.~Mollerach,
  Phys.\ Rev.\ D {\bf 42}, 313 (1990).
  
\bibitem{Linde:1996gt} 
  A.~D.~Linde and V.~F.~Mukhanov,
  Phys.\ Rev.\ D {\bf 56}, 535 (1997)
  [astro-ph/9610219].

\bibitem{Bartolo:2003jx}
  N.~Bartolo, S.~Matarrese and A.~Riotto,
  Phys.\ Rev.\ D {\bf 69} (2004) 043503
  [hep-ph/0309033].
  
  
  

\bibitem{Langlois:2004px}
  D.~Langlois and F.~Vernizzi,
  JCAP {\bf 0501} (2005) 002
  [astro-ph/0409684].
  
\bibitem{Battefeld:2014aea}
  T.~Battefeld and R.~C.~Freitas,
  JCAP {\bf 1409} (2014) 029
  [arXiv:1405.7969 [astro-ph.CO]].
  
\bibitem{Cespedes:2013rda}
  S.~CŽspedes and G.~A.~Palma,
  JCAP {\bf 1310} (2013) 051
  [arXiv:1303.4703 [hep-th]].
  
\bibitem{Noumi:2012vr}
  T.~Noumi, M.~Yamaguchi and D.~Yokoyama,
  JHEP {\bf 1306} (2013) 051
  [arXiv:1211.1624 [hep-th]].
  
\bibitem{Gao:2012uq}
  X.~Gao, D.~Langlois and S.~Mizuno,
  JCAP {\bf 1210} (2012) 040
  [arXiv:1205.5275 [hep-th]].
  
\bibitem{Achucarro:2012yr}
  A.~Achucarro, V.~Atal, S.~Cespedes, J.~O.~Gong, G.~A.~Palma and S.~P.~Patil,
  Phys.\ Rev.\ D {\bf 86} (2012) 121301
  [arXiv:1205.0710 [hep-th]].
  
\bibitem{Achucarro:2012sm}
  A.~Achucarro, J.~O.~Gong, S.~Hardeman, G.~A.~Palma and S.~P.~Patil,
  JHEP {\bf 1205} (2012) 066
  [arXiv:1201.6342 [hep-th]].
  
\bibitem{Cespedes:2012hu}
  S.~Cespedes, V.~Atal and G.~A.~Palma,
  JCAP {\bf 1205} (2012) 008
  [arXiv:1201.4848 [hep-th]].
  

  
\bibitem{Jackson:2010cw} 
  M.~G.~Jackson and K.~Schalm,
  Phys.\ Rev.\ Lett.\  {\bf 108}, 111301 (2012)
  [arXiv:1007.0185 [hep-th]].
  
\bibitem{future} 
  T.~Basse, J.~Hamann, S.~Hannestad and Y.~Y.~Y.~Wong,
  arXiv:1409.3469 [astro-ph.CO].
   
   \bibitem{fine}
 P.~A.~R.~Ade {\it et al.}  [Planck Collaboration],
  arXiv:1502.02114 [astro-ph.CO].
  
  
  \bibitem{selfrep} 
  A.~D.~Linde,
  Phys.\ Lett.\ B {\bf 175}, 395 (1986).
  
    \bibitem{station} 
  A.~D.~Linde and A.~Mezhlumian,
  Phys.\ Lett.\ B {\bf 307}, 25 (1993)
  [gr-qc/9304015].
  
  \bibitem{bigstation} 
  A.~D.~Linde, D.~A.~Linde and A.~Mezhlumian,
  Phys.\ Rev.\ D {\bf 49}, 1783 (1994)
  [gr-qc/9306035].
  
\bibitem{Goncharov:1987ir}
  A.~S.~Goncharov, A.~D.~Linde and V.~F.~Mukhanov,
  Int.\ J.\ Mod.\ Phys.\ A {\bf 2}, 561 (1987).
  
  
    \bibitem{bayes} 
  T.~Giannantonio and E.~Komatsu,
  arXiv:1407.4291 [astro-ph.CO].
  \bibitem{really?} 
  W.~H.~Kinney and K.~Freese,
  arXiv:1404.4614 [astro-ph.CO].
  
    \bibitem{<30} 
  R.~Easther and H.~Peiris,
  JCAP {\bf 0609}, 010 (2006)
  [astro-ph/0604214].

  
  
  \bibitem{pbh} 
  K.~Kohri, D.~H.~Lyth and A.~Melchiorri,
  JCAP {\bf 0804}, 038 (2008)
  [arXiv:0711.5006 [hep-ph]].
\bibitem{reconstruct} 
  J.~E.~Lidsey, A.~R.~Liddle, E.~W.~Kolb, E.~J.~Copeland, T.~Barreiro and M.~Abney,
  Rev.\ Mod.\ Phys.\  {\bf 69}, 373 (1997)
  [astro-ph/9508078].
  
  \bibitem{probe21} 
  H.~Shimabukuro, K.~Ichiki, S.~Inoue and S.~Yokoyama,
  Phys.\ Rev.\ D {\bf 90}, 083003 (2014)
  [arXiv:1403.1605 [astro-ph.CO]].
  

  
\bibitem{instep} 

 
  G.~Ballesteros, J.~A.~Casas and J.~R.~Espinosa,
  JCAP {\bf 0603}, 001 (2006)
  [hep-ph/0601134].

  G.~Ballesteros and J.~A.~Casas,
  arXiv:1406.3342 [astro-ph.CO].
  
\bibitem{bigsmall} 
  I.~Ben-Dayan and R.~Brustein,
  JCAP {\bf 1009}, 007 (2010)
  [arXiv:0907.2384 [astro-ph.CO]].  

  \bibitem{multisin} 
  T.~Kobayashi and F.~Takahashi,
  JCAP {\bf 1101}, 026 (2011)
  [arXiv:1011.3988 [astro-ph.CO]].
  
  \bibitem{just1} 
  Q.~E.~Minor and M.~Kaplinghat,
  arXiv:1411.0689 [astro-ph.CO].

\bibitem{falsefat} 
  M.~Peloso, L.~Sorbo and G.~Tasinato,
  JCAP {\bf 1406}, 040 (2014)
  [arXiv:1401.7136 [astro-ph.CO]].
  
    \bibitem{curv} 
  M.~S.~Sloth,
  Phys.\ Rev.\ D {\bf 90}, 063511 (2014)
  [arXiv:1403.8051 [hep-ph]].
    \bibitem{accu} 
  E.~D.~Stewart and D.~H.~Lyth,
  Phys.\ Lett.\ B {\bf 302}, 171 (1993)
  [gr-qc/9302019].
  \bibitem{veryhigh} 
  C.~P.~Burgess, J.~M.~Cline, F.~Lemieux and R.~Holman,
  JHEP {\bf 0302}, 048 (2003)
  [hep-th/0210233].  
  
  \bibitem{fastroll} 
  A.~D.~Linde,
  JHEP {\bf 0111}, 052 (2001)
  [hep-th/0110195].
  
  \bibitem{imaginary} 
  E.~J.~Weinberg and A.~q.~Wu,
  Phys.\ Rev.\ D {\bf 36}, 2474 (1987).
        \bibitem{chaotic} 
  A.~D.~Linde,
  Phys.\ Lett.\ B {\bf 129}, 177 (1983).

  \bibitem{hilltop} 
  L.~Boubekeur and D.~H.~Lyth,
  JCAP {\bf 0507}, 010 (2005)
  [hep-ph/0502047].
\bibitem{cole} 
  S.~R.~Coleman and E.~J.~Weinberg,
  Phys.\ Rev.\ D {\bf 7}, 1888 (1973).

\bibitem{dispar} 
  M.~Bastero-Gil, A.~Berera and B.~M.~Jackson,
  JCAP {\bf 1107}, 010 (2011)
  [arXiv:1003.5636 [hep-ph]].

  \bibitem{inverted} 
  A.~H.~Guth and S.~Y.~Pi,
  Phys.\ Rev.\ Lett.\  {\bf 49}, 1110 (1982).

\bibitem{boy1} 
  D.~Boyanovsky and H.~J.~de Vega,
  Phys.\ Rev.\ D {\bf 47}, 2343 (1993)
  [hep-th/9211044].

\bibitem{boy2} 
  D.~Boyanovsky, D.~s.~Lee and A.~Singh,
  Phys.\ Rev.\ D {\bf 48}, 800 (1993)
  [hep-th/9212083].

  \bibitem{preheat} 
  G.~N.~Felder, J.~Garcia-Bellido, P.~B.~Greene, L.~Kofman, A.~D.~Linde and I.~Tkachev,
  Phys.\ Rev.\ Lett.\  {\bf 87}, 011601 (2001)
  [hep-ph/0012142].

  \bibitem{monomial} 
  K.~Enqvist and M.~Karciauskas,
  JCAP {\bf 1402}, 034 (2014)
  [arXiv:1312.5944 [astro-ph.CO]].
  
    \bibitem{fullcalc} 
  T.~Markkanen and A.~Tranberg,
  JCAP {\bf 1211}, 027 (2012)
  [arXiv:1207.2179 [gr-qc]].
\bibitem{Candelas:1975du} 
  P.~Candelas and D.~J.~Raine,
  Phys.\ Rev.\ D {\bf 12}, 965 (1975).
  
  
\bibitem{Miao:2006pn} 
  S.~P.~Miao and R.~P.~Woodard,
  Phys.\ Rev.\ D {\bf 74}, 044019 (2006)
  [gr-qc/0602110].
  
\bibitem{Garbrecht:2006df} 
  B.~Garbrecht,
  Phys.\ Rev.\ D {\bf 74}, 043507 (2006)
  [hep-th/0604166].
  
  
\bibitem{Parker:1999td}
  L.~Parker and A.~Raval,
  Phys.\ Rev.\ D {\bf 60} (1999) 063512
   [Erratum-ibid.\ D {\bf 67} (2003) 029901]
  [gr-qc/9905031].
  
  
  
  
  
\bibitem{Giddings:2010nc} 
  S.~B.~Giddings and M.~S.~Sloth,
  JCAP {\bf 1101}, 023 (2011)
  [arXiv:1005.1056 [hep-th]].
  
\bibitem{Giddings:2010ui}
  S.~B.~Giddings and M.~S.~Sloth,
  JCAP {\bf 1007} (2010) 015
  [arXiv:1005.3287 [hep-th]].
   
\bibitem{Giddings:2011ze} 
  S.~B.~Giddings and M.~S.~Sloth,
  Phys.\ Rev.\ D {\bf 86}, 083538 (2012)
  [arXiv:1109.1000 [hep-th]].
  
  
\bibitem{Giddings:2011zd}
  S.~B.~Giddings and M.~S.~Sloth,
  Phys.\ Rev.\ D {\bf 84} (2011) 063528
  [arXiv:1104.0002 [hep-th]].
  
  
\bibitem{Dai:2013kra}
  L.~Dai, D.~Jeong and M.~Kamionkowski,
  Phys.\ Rev.\ D {\bf 88} (2013) 4,  043507
  [arXiv:1306.3985 [astro-ph.CO]].






\bibitem{Vilenkin:1982sg}
  A.~Vilenkin,
  Nucl.\ Phys.\ B {\bf 226}, 504 (1983);
  A.~Vilenkin,
  Nucl.\ Phys.\ B {\bf 226} (1983) 527.
  
\bibitem{Starobinsky:1994bd}
  A.~A.~Starobinsky and J.~Yokoyama,
  Phys.\ Rev.\ D {\bf 50} (1994) 6357
  [astro-ph/9407016].
  
\bibitem{Mukhanov:1996ak} 
  V.~F.~Mukhanov, L.~R.~W.~Abramo and R.~H.~Brandenberger,
  Phys.\ Rev.\ Lett.\  {\bf 78}, 1624 (1997)
  [gr-qc/9609026].
  L.~R.~W.~Abramo, R.~H.~Brandenberger and V.~F.~Mukhanov,
  Phys.\ Rev.\ D {\bf 56}, 3248 (1997)
  [gr-qc/9704037].

  
\bibitem{Abramo:1998hi}
  L.~R.~W.~Abramo and R.~P.~Woodard,
  Phys.\ Rev.\ D {\bf 60} (1999) 044010
  [astro-ph/9811430].
  L.~R.~W.~Abramo and R.~P.~Woodard,
  Phys.\ Rev.\ D {\bf 60}, 044011 (1999)
  [astro-ph/9811431].
  L.~R.~Abramo and R.~P.~Woodard,
  Phys.\ Rev.\ D {\bf 65}, 063515 (2002)
  [astro-ph/0109272].
  
  
 \bibitem{Losic:2005vg}
  B.~Losic and W.~G.~Unruh,
  Phys.\ Rev.\ D {\bf 72} (2005) 123510
  [gr-qc/0510078].
  B.~Losic and W.~G.~Unruh,
  Phys.\ Rev.\ Lett.\  {\bf 101} (2008) 111101
  [arXiv:0804.4296 [gr-qc]].
  
\bibitem{Sloth:2006az}
  M.~S.~Sloth,
  Nucl.\ Phys.\ B {\bf 748} (2006) 149
  [astro-ph/0604488].
  M.~S.~Sloth,
  Nucl.\ Phys.\ B {\bf 775} (2007) 78
  [hep-th/0612138].

  
\bibitem{Finelli:2001bn}
  F.~Finelli, G.~Marozzi, G.~P.~Vacca and G.~Venturi,
  Phys.\ Rev.\ D {\bf 65} (2002) 103521
  [gr-qc/0111035].
  F.~Finelli, G.~Marozzi, G.~P.~Vacca and G.~Venturi,
  Phys.\ Rev.\ D {\bf 69}, 123508 (2004)
  [gr-qc/0310086].
  P.~Martineau and R.~H.~Brandenberger,
  Phys.\ Rev.\ D {\bf 72} (2005) 023507
  [astro-ph/0505236].
  D.~Boyanovsky, H.~J.~de Vega and N.~G.~Sanchez,
  Phys.\ Rev.\ D {\bf 72} (2005) 103006
  [astro-ph/0507596].
  A.~Bilandzic and T.~Prokopec,
  Phys.\ Rev.\ D {\bf 76} (2007) 103507
  [arXiv:0704.1905 [astro-ph]].
  M.~van der Meulen and J.~Smit,
  JCAP {\bf 0711} (2007) 023
  [arXiv:0707.0842 [hep-th]].
  D.~Seery,
  JCAP {\bf 0802} (2008) 006
  [arXiv:0707.3378 [astro-ph]].
  D.~Seery,
  JCAP {\bf 0711} (2007) 025
  [arXiv:0707.3377 [astro-ph]].
  D.~H.~Lyth,
  JCAP {\bf 0712} (2007) 016
  [arXiv:0707.0361 [astro-ph]].
  E.~Dimastrogiovanni and N.~Bartolo,
  JCAP {\bf 0811} (2008) 016
  [arXiv:0807.2790 [astro-ph]].
  K.~Enqvist, S.~Nurmi, D.~Podolsky and G.~I.~Rigopoulos,
  JCAP {\bf 0804} (2008) 025
  [arXiv:0802.0395 [astro-ph]].
  N.~Bartolo, S.~Matarrese, M.~Pietroni, A.~Riotto and D.~Seery,
  JCAP {\bf 0801} (2008) 015
  [arXiv:0711.4263 [astro-ph]].
  N.~Afshordi and R.~H.~Brandenberger,
  Phys.\ Rev.\ D {\bf 63} (2001) 123505
  [gr-qc/0011075].
  T.~M.~Janssen, S.~P.~Miao, T.~Prokopec and R.~P.~Woodard,
  Class.\ Quant.\ Grav.\  {\bf 25} (2008) 245013
  [arXiv:0808.2449 [gr-qc]].
  P.~Adshead, R.~Easther and E.~A.~Lim,
  Phys.\ Rev.\ D {\bf 79} (2009) 063504
  [arXiv:0809.4008 [hep-th]].
  V.~Assassi, D.~Baumann and D.~Green,
  JHEP {\bf 1302}, 151 (2013)
  [arXiv:1210.7792 [hep-th]].
  S.~Nurmi, C.~T.~Byrnes and G.~Tasinato,
  JCAP {\bf 1306} (2013) 004
  [arXiv:1301.3128 [astro-ph.CO]].
  F.~Gautier and J.~Serreau,
  Phys.\ Lett.\ B {\bf 727}, 541 (2013)
  [arXiv:1305.5705 [hep-th]].
  V.~K.~Onemli,
  Phys.\ Rev.\ D {\bf 89}, no. 8, 083537 (2014)
  [arXiv:1312.6409 [astro-ph.CO]].
  T.~M.~Janssen, S.~P.~Miao, T.~Prokopec and R.~P.~Woodard,
  JCAP {\bf 0905} (2009) 003
  [arXiv:0904.1151 [gr-qc]].
  E.~T.~Akhmedov and P.~Burda,
  Phys.\ Rev.\ D {\bf 86} (2012) 044031
  [arXiv:1202.1202 [hep-th]].
  B.~Garbrecht, G.~Rigopoulos and Y.~Zhu,
  Phys.\ Rev.\ D {\bf 89} (2014) 6,  063506
  [arXiv:1310.0367 [hep-th]].
  C.~P.~Burgess, R.~Holman, G.~Tasinato and M.~Williams,
  arXiv:1408.5002 [hep-th].
  
  
\bibitem{Riotto:2008mv}
  A.~Riotto and M.~S.~Sloth,
  JCAP {\bf 0804} (2008) 030
  [arXiv:0801.1845 [hep-ph]].
  
\bibitem{Weinberg:2005vy}
  S.~Weinberg,
  Phys.\ Rev.\ D {\bf 72} (2005) 043514
  [hep-th/0506236].
  S.~Weinberg,
  Phys.\ Rev.\ D {\bf 74} (2006) 023508
  [hep-th/0605244].
  
  
  
   
  
\bibitem{Urakawa:2009gb}
  Y.~Urakawa and T.~Tanaka,
  Prog.\ Theor.\ Phys.\  {\bf 122} (2010) 1207
  [arXiv:0904.4415 [hep-th]].
  Y.~Urakawa and T.~Tanaka,
  Prog.\ Theor.\ Phys.\  {\bf 122} (2009) 779
  [arXiv:0902.3209 [hep-th]].
  
\bibitem{Byrnes:2010yc}
  C.~T.~Byrnes, M.~Gerstenlauer, A.~Hebecker, S.~Nurmi and G.~Tasinato,
  JCAP {\bf 1008} (2010) 006
  [arXiv:1005.3307 [hep-th]].
  M.~Gerstenlauer, A.~Hebecker and G.~Tasinato,
  JCAP {\bf 1106} (2011) 021
  [arXiv:1102.0560 [astro-ph.CO]].
  
\bibitem{Seery:2009hs}
  D.~Seery,
  JCAP {\bf 0905} (2009) 021
  [arXiv:0903.2788 [astro-ph.CO]].
 
  
  
\bibitem{Burgess:2009bs}
  C.~P.~Burgess, L.~Leblond, R.~Holman and S.~Shandera,
  JCAP {\bf 1003} (2010) 033
  [arXiv:0912.1608 [hep-th]].
  C.~P.~Burgess, R.~Holman, L.~Leblond and S.~Shandera,
  JCAP {\bf 1010} (2010) 017
  [arXiv:1005.3551 [hep-th]].

\bibitem{vilko} 
  G.~A.~Vilkovisky,
  Nucl.\ Phys.\ B {\bf 234}, 125 (1984).  
    \bibitem{mutated} 
  E.~D.~Stewart,
  Phys.\ Lett.\ B {\bf 345}, 414 (1995)
  [astro-ph/9407040].
  

   \bibitem{rolling} 
  B.~Ratra and P.~J.~E.~Peebles,
  Phys.\ Rev.\ D {\bf 37}, 3406 (1988).
  
   \bibitem{brane} 
  G.~Shiu and S.~H.~H.~Tye,
  Phys.\ Lett.\ B {\bf 516}, 421 (2001)
  [hep-th/0106274].
  

  
 
  
  \bibitem{susyinf} 
  G.~R.~Dvali, Q.~Shafi and R.~K.~Schaefer,
  Phys.\ Rev.\ Lett.\  {\bf 73}, 1886 (1994)
  [hep-ph/9406319].
\bibitem{Buchmuller:2014epa} 
  W.~Buchmüller, V.~Domcke, K.~Kamada and K.~Schmitz,
  JCAP {\bf 1407}, 054 (2014)
  [arXiv:1404.1832 [hep-ph]].
   
\bibitem{Langlois:2004nn}
  D.~Langlois and F.~Vernizzi,
  Phys.\ Rev.\ D {\bf 70} (2004) 063522
  [astro-ph/0403258].
  
  
\bibitem{Kinney:2012ik} 
  W.~H.~Kinney, A.~M.~Dizgah, B.~A.~Powell and A.~Riotto,
  Phys.\ Rev.\ D {\bf 86}, 023527 (2012)
  [arXiv:1203.0693 [astro-ph.CO]].

\bibitem{Fonseca:2012cj}
  J.~Fonseca and D.~Wands,
  JCAP {\bf 1206} (2012) 028
  [arXiv:1204.3443 [astro-ph.CO]].

\bibitem{Enqvist:2013paa}
  K.~Enqvist and T.~Takahashi,
  JCAP {\bf 1310} (2013) 034
  [arXiv:1306.5958 [astro-ph.CO]].

\bibitem{Byrnes:2014xua} 
  C.~T.~Byrnes, M.~Corts and A.~R.~Liddle,
  arXiv:1403.4591 [astro-ph.CO].
\bibitem{Takahashi:2013tj}
  F.~Takahashi,
  JCAP {\bf 1306} (2013) 013
  [arXiv:1301.2834, arXiv:1301.2834 [astro-ph.CO]].
  
\bibitem{Peloso:2014oza}
  M.~Peloso, L.~Sorbo and G.~Tasinato,
  JCAP {\bf 1406} (2014) 040
  [arXiv:1401.7136 [astro-ph.CO]].

\bibitem{Enqvist:2014nsa}
  K.~Enqvist, D.~J.~Mulryne and S.~Nurmi,
  arXiv:1412.5973 [astro-ph.CO].
\bibitem{Ferreira:2014zia}
  R.~Z.~Ferreira and M.~S.~Sloth,
  JHEP {\bf 1412} (2014) 139
  [arXiv:1409.5799 [hep-ph]].
    \bibitem{2loop} 
  C.~Ford and D.~R.~T.~Jones,
  Phys.\ Lett.\ B {\bf 274}, 409 (1992)
  [Erratum-ibid.\ B {\bf 285}, 399 (1992)].
  


  
 
  

  

  



  

  



  
  

  

  
   
 
  
 
  

  
\end{thebibliography}
\end{document}